\documentclass[preprint,12pt,authoryear]{elsarticle}
\makeatletter
\def\ps@pprintTitle{%
  \let\@oddhead\@empty
  \let\@evenhead\@empty
  \def\@oddfoot{\vspace*{6cm}\hfill
    Preprint submitted to JMPS\hfill\thepage}%
  \let\@evenfoot\@oddfoot}
\makeatother

\usepackage{graphicx}
\usepackage{amssymb}
\usepackage{amsmath}
\usepackage{amsthm}
\usepackage{mathtools}
\usepackage{float}
\usepackage{xcolor}
\usepackage{setspace}
\usepackage{multirow}  
\usepackage{makecell}
\usepackage{lineno}
\usepackage{geometry}
\journal{JMPS}
\usepackage[colorlinks=true,
            citecolor=blue,
            linkcolor=blue,
            urlcolor=blue]{hyperref}
\begin{document}

\begin{frontmatter}
\title{Effect of discreteness on domain wall stability in a plate coupled to a foundation of bistable elements}

\author[upenn]{Dengge Jin}
\author[upenn]{Samuele Ferracin}
\author[UF]{Vincent Tournat}
\author[upenn]{Jordan R. Raney\corref{cor1}}
\cortext[cor1]{Corresponding author}\ead{raney@seas.upenn.edu}

\affiliation[upenn]{organization={Department of Mechanical Engineering and Applied Mechanics, University of Pennsylvania},
             city={Philadelphia},
             postcode={19104},
             state={PA},
             country={USA}}
\affiliation[UF]{organization={Laboratoire d'Acoustique de l'Université du Mans (LAUM), Le Mans Université},
             city={CUMR},
             postcode={6613},
             country={France}}

\begin{abstract}
Surfaces and structures capable of multiple stable configurations have attracted growing interest for on-demand shape morphing. In this work, we consider an elastic compliant plate coupled to a two-dimensional foundation comprising an array of bistable elements, a system that can form and retain complex continuous morphologies without sustained actuation via creation of stable domain walls separating regions in different stable states. These domain walls exhibit three distinct behaviors: expansion, shrinking, and metastable pinning. These arise from two limits of foundation discreteness. In the continuum limit, where bistable units are strongly coupled, domain walls undergo global phase transitions analogous to first-order phase transitions. In the anti-continuum limit, discreteness introduces Peierls–Nabarro–type energy modulations that lead to metastable pinning.
To quantify these behaviors and the transition between the two limits, we develop a reduced-order model that captures the total potential energy of configurations with domain walls and validate it using finite element analysis (FEA). For axisymmetric domain walls, the model yields phase diagrams identifying the regimes of expansion, shrinking, and pinning as functions of bistable-potential asymmetry, relative foundation discreteness, and domain-wall size. We then extend the analysis to non-axisymmetric geometries and establish local geometric criteria that predict the stability of convex and concave polygonal domain walls, confirmed by simulations. Together, these results clarify how discreteness enables stability through energy-landscape modulation, provide predictive design rules for multistable reconfigurable surfaces, and offer insights into domain-wall stability more generally in elastically coupled multistable metamaterials.
\end{abstract}


 \begin{highlights}
 \item Plates coupled to bistable foundations can form domain walls that expand, shrink, or become pinned.
 
\item These behaviors depend on bistable asymmetry, discreteness, and domain-wall size.

\item An energetics model predicts phase diagrams for expansion, shrinking, and pinning.

\item Stability of general shapes follows from axisymmetric domain-wall stability.

\item Results give design rules for multistable reconfigurable surface.

 \end{highlights}


\end{frontmatter}

\section{Introduction}
Surface morphology, characterized by features such as curvature and pattern periodicity, plays a crucial role in governing physical properties across a wide range of length scales and domains, including mechanics (\cite{cerda2003geometry, bhushan2013principles}), acoustics~(\cite{chen2010study, Hussein2014}), optics~(\cite{hendrikx2018re, shen2020high}), and fluid dynamics~(\cite{barbarino2011review, peretz2020underactuated}). This strong connection between geometric configuration and functional response has motivated growing interest in reconfigurable surfaces and structures that allow on-demand tuning of shape and, consequently, functionality during operation~(\cite{barbarino2011review, dai_multi-stable_2013, cui_highly_2015, rafsanjani_propagation_2019, yang_emergent_2020, li_theory_2020, udani_programmable_2021, li_reconfiguration_2021, zhang_concept_2022, liu_multistable_2022, boston_spanwise_2022, rahman_shape-retaining_2024, shen_curvature_2024, wu_surface_2025, liu_corrugated_2025}). While most conventional compliant surfaces require sustained external actuation to maintain a deformed configuration, mechanical bistability~(\cite{santer_compliant_2008, chi_bistable_2022}), characterized by non-convex potential energy landscapes typically represented by a double-well profile, enables passive retention of new stable configurations without continuous energy input. Leveraging this principle, recent studies have explored reconfigurable surfaces formed by coupling compliant structures with arrays of bistable elements. In this work, we focus on a system comprising an elastic compliant plate coupled to a two-dimensional array of bistable structural elements. Previous work (\cite{zhang_concept_2022}) has experimentally shown that a similar system can realize a wide range of stable, curved surface shapes. Compared to purely discrete reconfigurable architectures, the presence of the plate provides a continuous surface, broadening potential application opportunities. Apart from this application interest, the plate naturally provides long-range elastic interaction between neighboring bistable elements, distinguishing this system from classical nearest-neighbor models such as the $\phi^4$ lattice~(\cite{kevrekidis_dynamics_2000}) and from many existing multistable mechanical metamaterial arrays~(\cite{nadkarni_universal_2016, raney_stable_2016, jin2020guided, zhou_cooperative_2023, janbaz_diffusive_2024, jiao2024phase}).

The multiple stable configurations in such a system emerge from the collective interplay between the compliance of the plate and the bistability of the foundation units. In addition to the effect of the bistable units on the curvature of the plate, the plate itself causes coupling between adjacent bistable units, and thus the collective behavior is not simply determined by the bistability of isolated units. This could lead to a rich set of possible nonlinear effects. When the coupling between adjacent units is relatively weak, individual elements behave nearly independently (\cite{zhang_concept_2022}) (as also shown in other systems, e.g., Refs.~\cite{chirikjian_binary_1994, santer_compliant_2008, cui_highly_2015, udani_programmable_2021, tahidul_haque_reprogrammable_2024}), allowing the array to access a large number of metastable configurations. In the anti-continuum limit, where coupling is negligible compared to the unit onsite stiffness, local actuation induces only local transitions, and the number of stable states scales as $2^N$(\cite{chirikjian_binary_1994}), with $N$ the number of units. At the opposite extreme (continuum limit), the neighboring elements are strongly elastically coupled, and a localized snap-through induces a dynamic dissipative or diffusive global phase transition with the domain wall which separates two different stable regions propagating through the whole system (\cite{nadkarni_universal_2016, raney_stable_2016, jin2020guided, zhou_cooperative_2023, janbaz_diffusive_2024, jiao2024phase}). This collective behavior is analogous to a first-order phase transition (\cite{oxtoby_nucleation_1998, binder_theory_1987,jiao2024phase}), in which the entire medium converts from one phase to the other. Due to the spontaneous moving domain wall, such systems remain effectively bistable rather than multistable. In this work, we are interested in understanding the transition between these two extremes. Characterizing this transition regime is essential for the rational design of multistable reconfigurable surfaces and for enabling controlled shifts between bistability and multistability, yet systematic studies of this regime in such a system remain limited.

The main goals of this work are twofold. First, we study the transition from strongly-coupled global phase transition behavior (continuum limit) to weakly-coupled multistable behavior (anti-continuum limit) in the plate-bistable foundation system, and we establish criteria for the system parameters necessary to achieve multistability. Inspired by the similarity between the domain-wall motion in this system and the dislocation or phase boundary motion in crystals, we find that the multistability of the domain wall originates from the relative discreteness of the foundation, which leads to periodic potential energy modulation and metastable domain-wall pinning analogous to the Peierls-Nabarro (P-N) effect (\cite{kevrekidis_dynamics_2000, truskinovsky_peierls-nabarro_2003}). The discreteness can be quantified by the ratio between the lattice spacing of the bistable foundation units and the characteristic width of the domain wall. This discreteness ratio reflects the competition between the bistability of individual foundation units and the cooperative interactions mediated by the plate. To quantitatively assess how discreteness reshapes the energy landscape, we use collective coordinate approximation (\cite{gervais_extended_1975, takyi_collective_2016}) and develop an analytical reduced-order model (ROM) that applies consistently in both the continuum and discrete regimes. Second, within the multistable regime, we identify which domain-wall geometries remain stable. We begin with simple axisymmetric domain walls. Our model shows that metastability arises only within a certain domain-wall size range, reflecting the competition between discreteness effects (P-N pinning) and the global energetic drive for phase transition. We then establish a local–global stability criterion to infer the stability of general polygonal domain walls. Both the axisymmetric and polygonal domain-wall results from our model are validated against Finite-Element Simulation (FEA).


The remainder of the paper is organized as follows. Section \ref{section 2} introduces the plate–bistable foundation system and illustrates three representative domain-wall behaviors: expansion, shrinking, and metastable pinning. It also provides a qualitative explanation of the underlying mechanisms that give rise to these behaviors. Section \ref{section 3} develops the reduced-order model (ROM) and derives the potential energy for both the continuous and discrete foundation cases. Section \ref{section 4} employs the ROM to quantify the critical nucleation radius $r_{\text{cr}}$ which separate expansion from shrinking, and to analyze domain-wall dynamics in the continuous regime, as well as the effect of discreteness $\gamma$ on stability in the discrete regime. Section \ref{section 5} extends the stability analysis to general polygonal domain walls. Section \ref{section 6} concludes with a summary and discussion.

\section{Expansion, shrinking, or metastable pinning of a domain wall}\label{section 2}

\begin {figure}[H]
  \centering
  \includegraphics[width=1.0\linewidth]{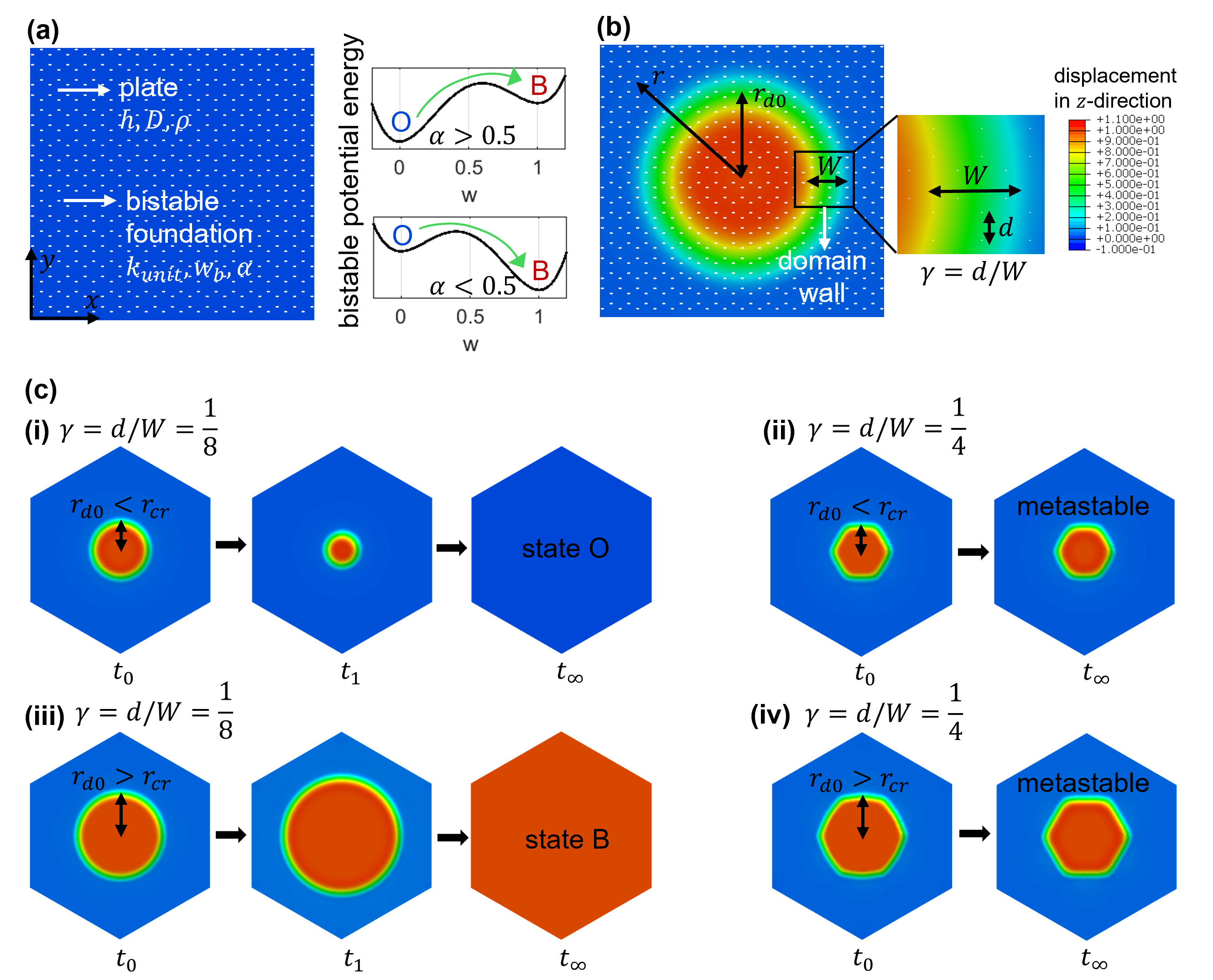}  
  \caption{(a) Plate-bistable foundation system in original configuration. The bistable units are arranged in a hexagonal pattern. Their stable states are indicated by O (with displacement in z-direction $w=0$) and B ($w=w_b$). (b)~Displacement boundary condition (perpendicular to the paper) is applied to a circular region, causing the interior region to deform from the original state O to the other stable B, forming a domain wall (in green) with radius $r_{d0}$ and width $W$. The insert shows the edge size of the hexagonal pattern is $d$. The relative density of the foundation is defined as $\gamma=d/W$. (c) FEA results: (i) and (iii) the domain wall moves when the foundation density $\gamma$ is relatively small ($\gamma=1/8$). It shrinks when the initial domain wall size $r_{d0}$ is smaller than a critical size $r_{cr}$ and expands when $r_{d0}>r_{cr}$; (ii) and (iv) the domain wall stabilizes and stays near its initial radius when the foundation density is relatively large with $\gamma=1/4$. All the FEA results in this plot are with parameters: $h=1, D=1, \rho=0.1, k=1, w_b=1, \alpha=0.48$ and the corresponding characteristic length is $L=1$. All the dimensions shown here are dimensionless, normalized by $L$ if in $x-y$ plane and by $w_b$ in $z$ direction. Moderate damping is used to obtain the metastable configurations.}
  \label{fig1}
\end{figure}

We consider an elastic thin plate with thickness $h$, bending stiffness $D=\frac{2h^3E}{3(1-\nu^2)}$, and density $\rho$ supported by a bistable foundation consisting of a two-dimensional array of bistable units, as shown in Fig.~\ref{fig1}(a).  To ensure isotropy, the bistable units are arranged in a hexagonal honeycomb pattern with an edge length $d$. A quartic polynomial is used to represent the bistable potential for a single unit:
\begin{equation}\label{bistable potential}
U_{\text {unit}}(w) = k_{\text {unit}} \left [ \frac{1}{4}w^4-\frac{1}{3}(1+\alpha)w_bw^3 +\frac{1}{2}\alpha w_b^2 w^2\right ]
\end{equation}
where $k_{\text{unit}}$ is the stiffness, $w$ is the out-of-plane displacement of the bistable unit, $w_b$ is the displacement from the initial stable equilibrium state $O$ to the another stable equilibrium state $B$, and $\alpha$ is a parameter that characterizes the asymmetry of the bistable potential. For $\alpha>0.5$, $U_{\text{unit}}(w=w_o=0)<U_{\text{unit}}(w=w_b)$ and vice versa. According to the geometry of the foundation pattern the effective stiffness per unit area for the foundation is defined as $k = \frac{4}{3\sqrt{3}} k_{\text{unit}}/d^2$.

Now we consider a scenario where a circular region of the foundation is imposed (actuated) via displacement control to deform from the initial state $O$ to the other stable state $B$, as shown in Fig.~\ref{fig1}(b). Due to the elastic coupling from the plate, a domain wall with radius $r_{d0}$ forms, separating the interior region B and the exterior region O. Within the domain wall, the bistable units are in an intermediate position between state $O$ and $B$. The domain wall, with a width $W$, can span multiple units. We define the discreteness parameter $\gamma$ by $\gamma=d/W$ to reflect the relative density of the bistable units inside the domain wall. Next, we release all the displacement boundary condition, and consider how the initial deformation of the plate (effectively the domain wall) evolves. 

We first examined the evolution of the domain wall using finite element simulations (\textit{FEA Software Abaqus}). The plate was modeled as a hexagon and discretized using a structured triangular mesh with S3R element (linear triangular shell element). A mesh-convergence study was performed to ensure that the mesh size was sufficiently small relative to the domain wall width. The choice of triangular elements preserves geometric compatibility with the underlying hexagonal pattern of the bistable foundation. Bistable foundation units were assigned at hexagonally patterned locations with spacing \(d\). Each foundation unit was represented by a SPRING1 element acting in the out-of-plane (\(z\)) direction. The bistable potential Eq.~(\ref{bistable potential}) and its corresponding force–displacement law were implemented by modifying the tabulated nonlinear spring response in the \texttt{.inp} file. To maintain consistency and enforce nondimensionality, all model parameters \((h, D, \rho, w_b)\) were set to unity, except for the unit spacing \(d\) and the bistable-element stiffness \(k_{\text{unit}}\). For each simulation case, \(d\) and \(k_{\text{unit}}\) were jointly varied to maintain a constant effective foundation stiffness, $k = \frac{4}{3\sqrt{3}}\,\frac{k_{\text{unit}}}{d^2}=1$,
thus ensuring that the characteristic length $L=\sqrt[4]{\frac{D}{k w_b^2}}$ of the system is unity and the domain-wall width \(W\) remains fixed (see Section~3 for the normalization and discussion on the domain wall width). Under this construction, the discreteness parameter \(\gamma = d/W\) becomes directly proportional to \(d\). The simulation procedure consisted of two steps. First, a prescribed displacement \(w = 1\) (normalized by \(w_b\)) was applied quasi-statically over a circular region of radius \(r_{d0}\) to nucleate a domain wall. In the second step, this displacement was removed, and the structure evolved under an implicit dynamic step. A moderate level of Rayleigh damping \((\alpha = \beta = 0.4)\) was introduced to suppress transient oscillations and allow the possible metastable configuration to be identified reliably. To map the full set of accessible stable configurations, we repeated this procedure for a wide range of initial radii \(r_{d0}\) at each value of \(\gamma\). Three different typical domain wall evolution behaviors are presented in Fig.~\ref{fig1}(c).

The evolution of a domain wall is governed by the system's potential energy $V$, which includes contributions from both the bending energy of the plate and the bistable elements.
The bending energy of the plate is localized within the domain wall, which scales approximately (in the first-order) with the perimeter $\sim r_d$ and is referred to as surface energy. The contribution from the bistable elements relates to the energy difference of the transition from states O to B, which approximately scales with $~r^2_d$ and is referred to as the bulk energy. For $\alpha>0.5$, as the domain wall expands, both the surface energy and the bulk energy increases, giving $V\sim O(r_d)+O(r_d^2)$. The domain wall therefore shrinks after the initial displacement is released. For $\alpha<0.5$, as the domain wall expands, the surface energy increases but the bulk energy decreases, giving $V \sim O(r_d)-O(r_d^2)$. Competition between the two terms produces a critical radius $r_{\text{cr}}$, below which the $O(r_d)$ term (surface energy) dominates and the domain wall is prone to shrink and above which the $-O(r^2_d)$ term (bulk energy) governs and the domain wall is prone to expand. These two typical behaviors for $\alpha<0.5$ are observed in FEA simulations in Fig.~\ref{fig1}(c1) and Fig.~\ref{fig1}(c3).

\begin{figure}[H]
  \centering
  \includegraphics[width=0.7\linewidth]{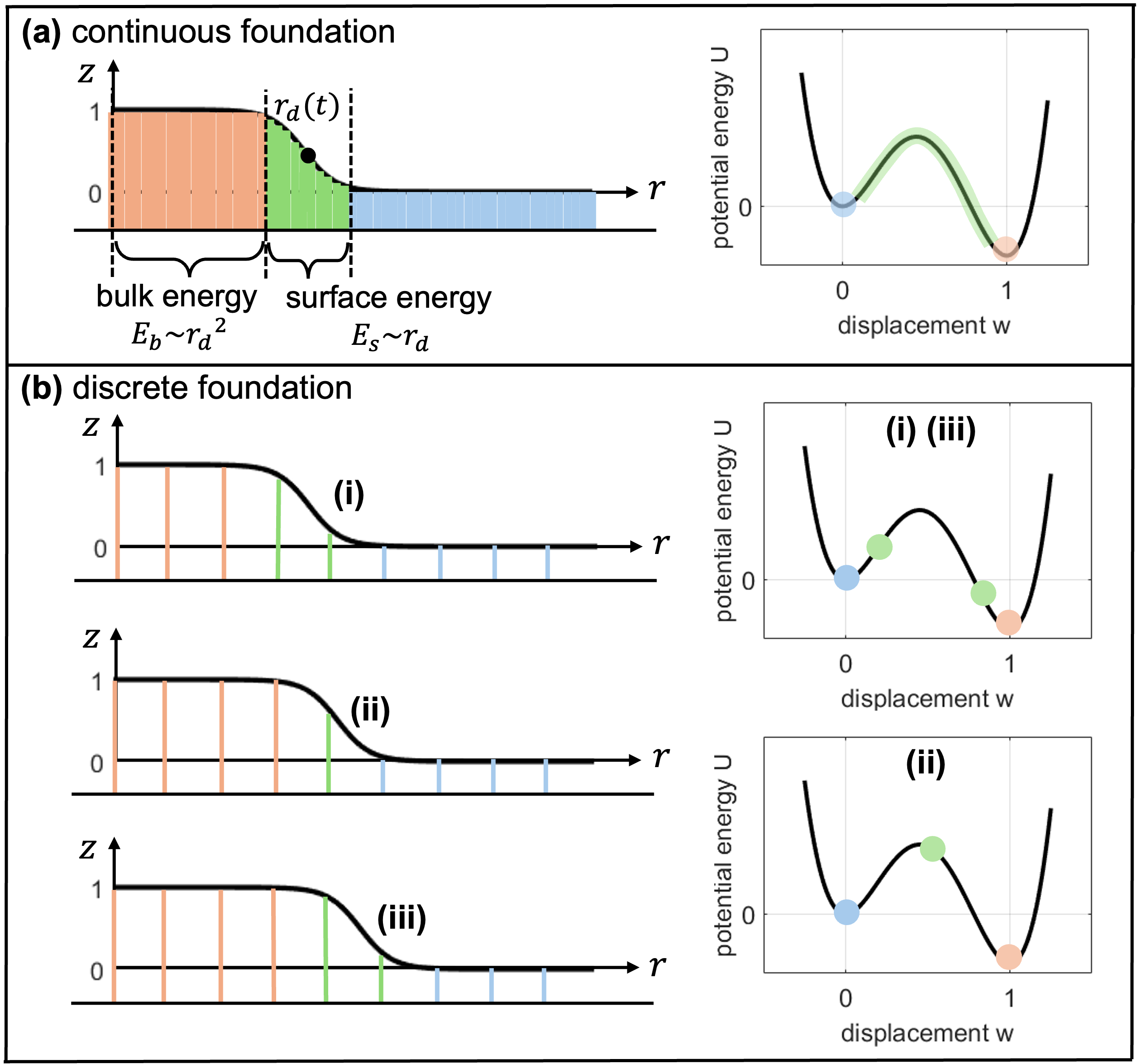}
  \caption{The energy distribution in the bistable foundation for (a) small $\gamma$ (continuous foundation) and (b) large $\gamma$ (discrete foundation), respectively. The red color denotes state O, blue denotes state B, and green denotes the domain wall (intermediate region between O and B). }
  \label{P-N effect}
\end{figure}

However, Fig.~\ref{fig1}(c2) and Fig.~\ref{fig1}(c4) show that when $\gamma=d/W$ is large enough, the domain wall can stabilize at a finite radius rather than expand or shrink. These metastable states correspond to local minima in the potential energy $V$, arising from the discreteness of the bistable foundation, an effect analogous to the P-N effect (\cite{kevrekidis_dynamics_2000, truskinovsky_peierls-nabarro_2003}). The P-N effect in our system can be illustrated by Fig.~\ref{P-N effect}, which conceptually shows the spatial distribution of foundation energy along the radial direction as the domain wall shifts. Considering a process where the domain wall moves from (i) to (ii) and then to (iii). When the foundation lattice size is negligible relative to the domain wall width ($\gamma$ is small), as in Fig.~\ref{P-N effect}(a), the energy distribution in the foundation remains nearly unchanged throughout the domain wall motion. However, when the lattice size is comparable to the domain wall width ($\gamma$ is large), as in Fig.~\ref{P-N effect}(b), the energy distributions at positions (i) and (iii) are similar due to lattice periodicity, but differ significantly at the intermediate position (ii) due to lattice discreteness. This results in a periodic variation in $V$ as the domain wall traverses the lattice, giving rise to local minima, i.e., the metastable states.

Although the initial loading area is circular, the final metastable domain wall conforms to the hexagonal symmetry of the foundation. This occurs because the system is more anisotropic at larger $\gamma$, and the domain wall naturally evolves toward a shape that minimizes energy with this anisotropy. An analogy can be found in crystal growth: when interfacial energy is anisotropic (as in snowflakes or hexagonal crystals), the equilibrium morphology forms facets rather than circles.

\section{Theoretical model}\label{section 3}

When the discreteness parameter $\gamma=d/W$ is sufficiently small, the force exerted on the plate by the foundation can be treated as continuously distributed force. We follow the classical Kirchhoff–Love theory for thin plates (\cite{timoshenko1959theory}) and the governing equation for the axisymmetric deformation $w(r,t)$ of the plate–bistable foundation system is~:
\begin{equation}
\rho h \frac{\partial^2 w}{\partial t^2} + D\left( \frac{\partial^4 w}{\partial r^4} + \frac{2}{r}\frac{\partial^3 w}{\partial r^3} - \frac{1}{r^2}\frac{\partial^2 w}{\partial r^2} + \frac{1}{r^3}\frac{\partial w}{\partial r} \right) + k w(w - \alpha w_b)(w - w_b) = 0
\end{equation}
where $r$ is the radial coordinate with its origin at the domain wall center, $k=\frac{4}{3\sqrt{3}}k_{\text{unit}}/d^2$ is the effective foundation stiffness as defined previously, and the last term is derived from Eq.~(\ref{bistable potential}) by taking the first derivative with respect to $w$. Based on this PDE, we define the characteristic time $T$ and length $L$, 
\begin{equation}
T = \sqrt{\frac{\rho h}{k w_b^2}}, \quad L = \sqrt[4]{\frac{D}{k w_b^2}}
\end{equation}
and nondimensional variables:
\begin{equation}
w = w_b \bar{w}, \quad t = T \bar{t}, \quad r = L \bar{r} ,
\end{equation}
leading to the normalized PDE,
\begin{equation}\label{PDE}
\frac{\partial^2 \bar{w}}{\partial \bar{t}^2} + \left( \frac{\partial^4 \bar{w}}{\partial \bar{r}^4} + \frac{2}{\bar{r}}\frac{\partial^3 \bar{w}}{\partial \bar{r}^3} - \frac{1}{\bar{r}^2}\frac{\partial^2 \bar{w}}{\partial \bar{r}^2} + \frac{1}{\bar{r}^3}\frac{\partial \bar{w}}{\partial \bar{r}} \right) + \bar{w}(\bar{w} - \alpha)(\bar{w} - 1) = 0 .
\end{equation}

\subsection{Reduced-order model}

\begin{figure}[H]
  \centering
  \includegraphics[width=0.7\linewidth]{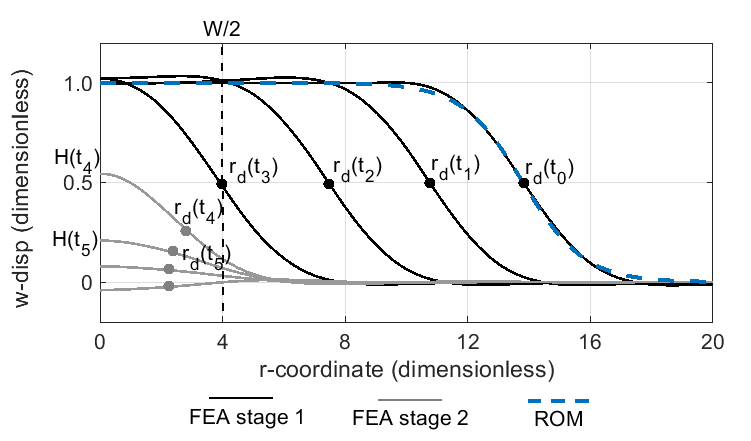}  
  \caption{Two-stage shrinking process of a domain wall. Dots mark the domain wall center positions. FEA results shown for \(\alpha=0.48\). w-displacement and r-coordinate are normalized by $w_b$ and characteristic length $L$, respectively.}
  \label{2stage}
\end{figure}

Eq.~\eqref{PDE} is closely related to the classical $\phi^4$ model, a nonlinear wave equation with a symmetric quartic bistable potential that has been extensively studied over the past several decades~(\cite{kevrekidis_dynamics_2000}). However, unlike the standard $\phi^4$ equation, Eq.~\eqref{PDE} incorporates a biharmonic operator arising from plate bending, which substantially complicates the analysis and has received only limited analytical treatment to date~(\cite{decker_kink_2021}). As a consequence, no closed-form solutions of Eq.~\eqref{PDE} are available. To overcome this difficulty, we adopt a reduced-order model (ROM) based on a hyperbolic tangent ansatz, Eq.~\eqref{tanh}, which is motivated by finite-element simulations (Fig.~\ref{2stage}) as well as by the known kink solutions of the classical $\phi^4$ model,
\begin{equation}\label{tanh}
w(r,t) = \frac{1}{2} H \left[ 1 - \tanh\!\left( \frac{r - r_d}{W/4} \right) \right],
\end{equation}  
where \( r_d \) is the domain wall center position,  \( H \) the height, and \( W \) the width. This general form Eq.~(\ref{tanh}) captures the plate shapes in two stages of evolution shown in Fig.~\ref{2stage}.

\textit{Stage 1} (\( r_d(t) > W/2 \)): The domain wall either expands or shrinks (black lines in Fig.~\ref{2stage}), with height \( H = w_b \). Although \( W/L \) in principle depends on the PDE parameter \( \alpha \) and the instantaneous position \( r_d(t) \), FEA results indicate that this dependence is weak and it can be approximated as a constant \( W/L \approx 8\). The factor \( W/4 \) in the \(\tanh\) argument ensures that for \( r \leq r_d(t) - W/2 \), \( w(r,t) \approx w_b \) (interior state B), and for \( r \geq r_d(t) + W/2 \), \( w(r,t) \approx 0 \) (exterior state O). The plate profile Eq.~(\ref{tanh}) reduces to:
\begin{equation}\label{tanh1}
w(r,t) = \frac{1}{2} w_b \left[ 1 - \tanh\!\left( \frac{r - r_d(t)}{W/4} \right) \right]
\end{equation}

\textit{Stage 2} (\( r_d(t) < W/2 \)): The domain wall is vanishing (gray lines in Fig.~\ref{2stage}). Both the height \( H(t) \) and width \( W(t) \) vary with time. Since \( r_d(t) = W(t)/2 \), the two independent, time-dependent parameters are \( H(t) \) and \( r_d(t) \) (or equivalently \( W(t) \)):
\begin{equation}\label{tanh2}
w(r,t) = \frac{1}{2} H(t) \left[ 1 - \tanh\!\left( \frac{r - r_d(t)}{r_d(t)/2} \right) \right].
\end{equation}
To study the stability and dynamics of the domain wall, we apply the collective coordinate method (\cite{gervais_extended_1975, takyi_collective_2016}). This approach assumes that the displacement field maintains a fixed form (here, a \(\tanh\) profile), while the system’s evolution is governed by a small set of time-dependent parameters (the collective coordinates). In our approximation, Stage~1 is characterized solely by \( r_d(t) \), whereas Stage~2 involves both \( r_d(t) \) and \( H(t) \). This reduced-order model (ROM) enables explicit derivation of an effective potential energy landscape \( U(r_d, H) \), from which key mechanical features, such as energy maxima and minima, can be identified. The ROM thus provides the foundation for quantitatively predicting domain wall evolution, stability, and reconfigurability in the plate--bistable foundation system.

\subsection{Potential energy for continuous foundation case}
We first consider the foundation in a continuous limit where $\gamma=d/W \ll 1$. In stage~1 where the domain wall either shrinks or expands, the elastic energy in the plate (\cite{timoshenko1959theory}) is given by~
\begin{equation}\label{eq: plate energy main 1}
\begin{aligned}
E_{p1} &= \frac{1}{2} D \iint \left[\frac{1}{r^2} \left( \frac{\partial w}{\partial r} \right)^2 +\left( \frac{\partial^2 w}{\partial r^2} \right)^2 \right] dA \\
& \approx \pi D w_b^2\int_{-\infty}^\infty \left[  \frac{4}{W^2} \text{sech}^4(\xi) \frac{1}{r_d} +\frac{16^2}{W^4} \text{sech}^4(\xi) \text{tanh}^2(\xi) r_d \right] \frac{W}{4} d\xi\\
& = \frac{\pi D w_b^2}{15 W^2} \left( 256 \frac{r_d}{W} + 20 \frac{W}{r_d} \right)
\end{aligned}
\end{equation}
where the deformation Eq.~(\ref{tanh1}) is used and Poisson ratio is taken as zero for simplicity without losing the key insights by the ROM. 

The elastic energy stored in the foundation is calculated as
\begin{equation}\label{eq:foundation energy main 1}
\begin{aligned}
E_{f1}
&= \int_{0}^{\infty} 2\pi r\, U\!\big(w(r)\big)\,dr \\
&= 2\pi k \!\int_{0}^{\infty}\!\!\left[ 
\tfrac{1}{4} w^4
-\tfrac{1}{3}(1+\alpha)w_b\,w^3
+\tfrac{1}{2}\alpha w_b^{2}w^{2}
\right] r\,dr \\[4pt]
& \approx   \frac{\pi k\, w_b^{4} W^{2}}{384}
\left[
   (2\alpha - 1)
   \;+\;
   4\,\frac{r_d}{W}
   \;+\;
   32\,(2\alpha - 1)\left(\frac{r_d}{W}\right)^{2}
\right]
\end{aligned}
\end{equation}
where $U(w)$ is the bistable potential Eq.~(\ref{bistable potential}) and the ansatz Eq.~(\ref{tanh1}) is used. To yield the above explicit expression, the change of variable $\frac{r-r_d}{W/4}=\xi$  was used and the integration limit $-\frac{r_d}{W/4}$ was well approximated by $-\infty$ due to the strong localization of $\text{sech}$ function. The derivation details are presented in \ref{append A}.

Next, we consider stage 2 where the domain wall disappears. The elastic energy in the plate in stage 2 follows from Eq.~(\ref{eq: plate energy main 1}) with $w_b$ replaced by $H(t)$ and $W(t)=2r_d(t)$:
\begin{equation}\label{eq: plate energy main 2}
E_{p2} = \frac{14\pi D}{5} \frac{H^2(t)}{{r_d(t)}^2}
\end{equation}

The elastic energy in the foundation is:
\begin{equation}\label{eq: foundation energy main 2}
\begin{aligned}
E_{f2} &= \int_0^\infty 2\pi r \, U(w(r)) \, dr \\
&= 2\pi  k \int_0^\infty \left [ \frac{1}{4}w^4-\frac{1}{3}(1+\alpha)w_bw^3 +\frac{1}{2}\alpha w_b^2  w^2\right ]   r dr \\
&\approx \frac{\pi k w_b^4}{96}  \left [ 11\,{\left ( \frac{H(t)}{w_b} \right )}^4-(18\alpha+18)\,{\left ( \frac{H(t)}{w_b} \right )}^3+36 \alpha\,{\left ( \frac{H(t)}{w_b} \right )}^2 \right ] {r_d(t)}^2
\end{aligned}
\end{equation}
which recovers Eq.~(\ref{eq:foundation energy main 1}) for $H = w_b$ and $r_d = W/2$.

Defining the characteristic energy based on the characteristic scales as
\begin{equation}
E = k w_b^4 L^2,
\end{equation}
the energies in Eqs.~(\ref{eq: plate energy main 1})-(\ref{eq: foundation energy main 2}) can be normalized. For \( r_d(t) > W/2 = 4 \), we obtain
\begin{align}
\label{normalized energy 11}
\overline{E}_{p1} = \frac{E_{p1}}{\pi E} &= \frac{1}{15W^2} \left( 256 \frac{r_d}{W} + 20 \frac{W}{r_d} \right), \\
\label{normalized energy 12}
\overline{E}_{f1} = \frac{E_{f1}}{\pi E} &= \frac{W^2}{384} \left[ (2\alpha - 1) + 4\frac{r_d}{W} + 32(2\alpha - 1)\left(\frac{r_d}{W}\right)^2 \right] ,
\end{align}
and for \( r_d(t) \leq 4 \), we have:
\begin{align}
\label{normalized energy 21}
\overline{E}_{p2} = \frac{E_{p2}}{\pi E} &= \frac{14}{5} \frac{H^2(t)}{r_d^2(t)}, \\
\label{normalized energy 22}
\overline{E}_{f2} = \frac{E_{f2}}{\pi E} &= \frac{1}{96} \left[ 11 H^4(t) - (18\alpha + 18) H^3(t) + 36\alpha H^2(t) \right] r_d^2(t).
\end{align}
In the above equations, \( r_d \) and \( W \) are normalized by the characteristic length \( L \), and \( H \) by \( w_b \). For simplicity, we omit the overbar notation for normalized quantities throughout the rest of the analysis.

The total potential energy \( V(r_d, H) \), consisting of the elastic energy in the plate and foundation, is given by:
\begin{equation}
\label{potential energy}
V(r_d, H) =
\begin{cases}
{E}_{p1} + {E}_{f1} = V_1(r_d), & \text{if } r_d > \frac{W}{2} = 4, \\[8pt]
{E}_{p2} + {E}_{f2} = V_2(r_d, H), & \text{if } r_d < \frac{W}{2} = 4.
\end{cases}
\end{equation}

As seen from Eqs.~(\ref{normalized energy 11}) and (\ref{normalized energy 12}), the potential energy in Stage 1, \( V_1(r_d) \), consists of three terms: a reciprocal term \( \sim 1/r_d \), a linear term \( \sim r_d \), and a quadratic term \( \sim r_d^2 \) (the constant term only affects the energy level but does not change the energy landscape). The reciprocal term originates from the azimuthal bending curvature, which is significant for small \( r_d \) but becomes negligible as \( r_d \) increases, since the deformation approaches a nearly flat arc and the azimuthal curvature vanishes in the large-\( r_d \) limit. The linear term, which has a positive coefficient, corresponds to the surface energy of the phase transition. The coefficient of the quadratic term depends on the parameter \( \alpha \): it is positive for \( \alpha > 0.5 \) and negative for \( \alpha < 0.5 \), reflecting the bulk energy gain or loss during the phase transition. These features from the reduced-order model are consistent with qualitative arguments given earlier.

\begin{figure}[H]
    \centering
    \includegraphics[width=1.0\linewidth]{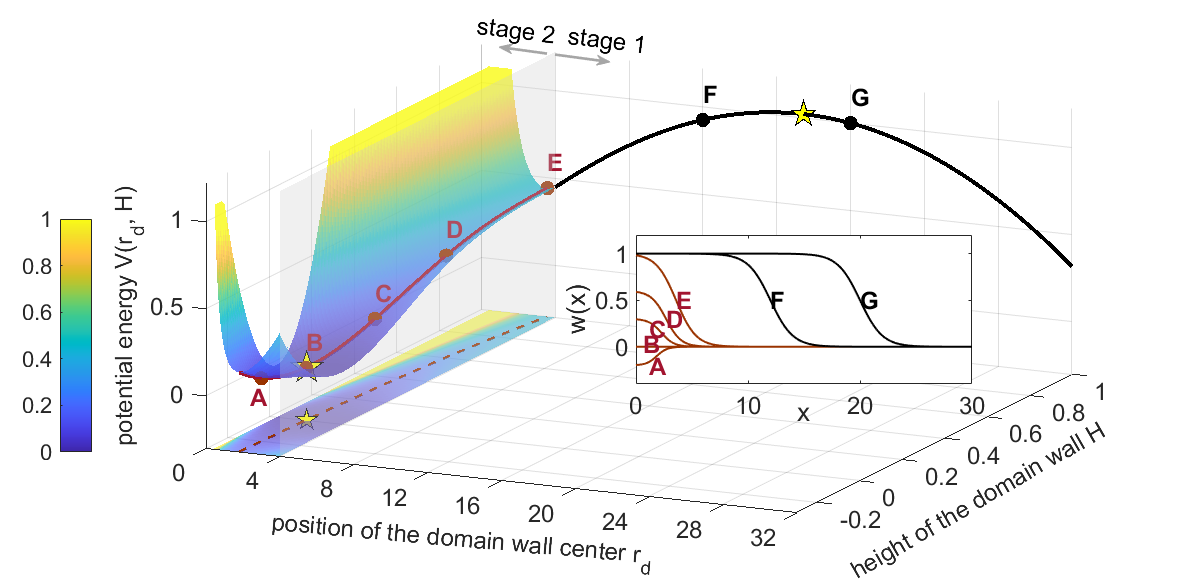}
    \caption{The potential energy $V(r_d,H)$ of the reduced order model (ROM) with $\alpha=0.48$. The two stars represent the global maximum in stage 1 (domain wall expanding or shrinking) and the local minimum in stage 2 (domain wall disappearing), respectively. The trajectory FEDCBA is a possible evolution of the domain wall. The insert shows the profiles of the plate with different domain wall positions.}
    \label{fig:continuous potential}
\end{figure}

From Eqs.~(\ref{normalized energy 11}), (\ref{normalized energy 12}), (\ref{normalized energy 21}), and (\ref{normalized energy 22}), it follows that the potential energy landscape \( V(r_d, H) \) is primarily governed by the bistable foundation parameter \( \alpha \). An illustrative case for \( \alpha = 0.48 \) is shown in Fig.~\ref{fig:continuous potential}.

In Stage 1, where the domain wall expands or shrinks, the potential energy depends only on the wall position $r_d$ and exhibits a global maximum at the \textit{critical nucleation size} $r_{cr}$. If a circular domain wall with $r_{d0} > r_{cr}$ (e.g., point G in Fig.~\ref{fig:continuous potential}) is created and released, the positive driving force $f = -\, dV/dr_d > 0$ drives it to expand until it reaches the system boundary, restoring the plate to a flat configuration. Conversely, if $r_{d0} < r_{cr}$ (e.g., point F), the negative driving force $f < 0$ causes the wall to contract toward point~E. When the evolution reaches point E, the system enters Stage 2, where the potential energy depends on both $r_d$ and $H$. A local minimum exists at point B, satisfying $\partial V/\partial r_d = 0$ and $\partial V/\partial H = 0$ with a positive definite Hessian. This minimum always occurs at $H = 0$, independent of $\alpha$, corresponding to the original flat state.

The trajectory in the $(r_d, H)$ space during Stage~2 depends on the incoming velocities $\dot r_d(t_E)$ and $\dot H(t_E)$. Fig.~\ref{fig:continuous potential} presents one possible path. In a conservative system, these trajectories form closed loops corresponding to periodic nonlinear oscillations. In dissipative systems, however, energy is gradually lost, and all trajectories ultimately settle into the stable minimum at point~B.

\subsection{Potential energy for discrete foundation case}
Now we consider the foundation in a more discrete case. Our objective is to investigate how the discreteness of the foundation influences the system’s potential energy $V$. As defined previously, with the hexagonal lattice edge size $d$ and the bistable unit stiffness $k_{\text{unit}}$, the effective bistable foundation stiffness per unit area is defined as  $k= \frac{4}{3\sqrt{3}} \frac{k_{unit}}{d^2}$ and the characteristic length of the system is defined as $L = \sqrt[4]{\frac{D}{k w_b^2}}$. To approximate the potential energy, when the foundation density is not far away from the continuous limit, we assume that the domain wall has a circular shape and the width of the domain wall is approximately $W=8L$, the same with the continuous limit. Therefore, the plate deformation still follows the same shape as described by Eq.~(\ref{tanh}). To quantify the discreteness of the foundation, we have defined the normalized edge size of the hexagon pattern as
\(\gamma=\frac{d}{W}\). The above assumptions are asymptotic as the results converges to the continuous case when $\gamma \rightarrow 0$. With these assumptions for deformation, the elastic energy due to plate deformation can be integrated in the same way with the continuous case, while the elastic energy in the foundation is obtained by summing the contributions from all bistable units:
\begin{equation}
E_f = \sum_{i=1}^{N} U_{unit}\left( w(r_i) \right) 
\end{equation}
where $N$ is the number of the bistable units in the foundation and $r_i$ is the radial position of the $i$th unit.

\section{Axisymmetric domain walls}\label{section 4}

In this section, we present the behavior of an axisymmetric domain wall predicted by the ROM in Section \ref{section 3} and provide quantitative comparison with FEA simulations. We first focus on the continuous limit in Section \ref{subsection 4.1} where the domain wall either shrinks or expands, and then we discuss the transition from the continuous limit to discrete foundation case and how the discreteness stabilizes the domain wall in Section \ref{subsection 4.2}. 

\subsection {Continuum limit} \label{subsection 4.1}
\subsubsection{Critical nucleation size}\label{section 4.1.1}
\begin{figure}[H]
  \centering
  \includegraphics[width=\linewidth]{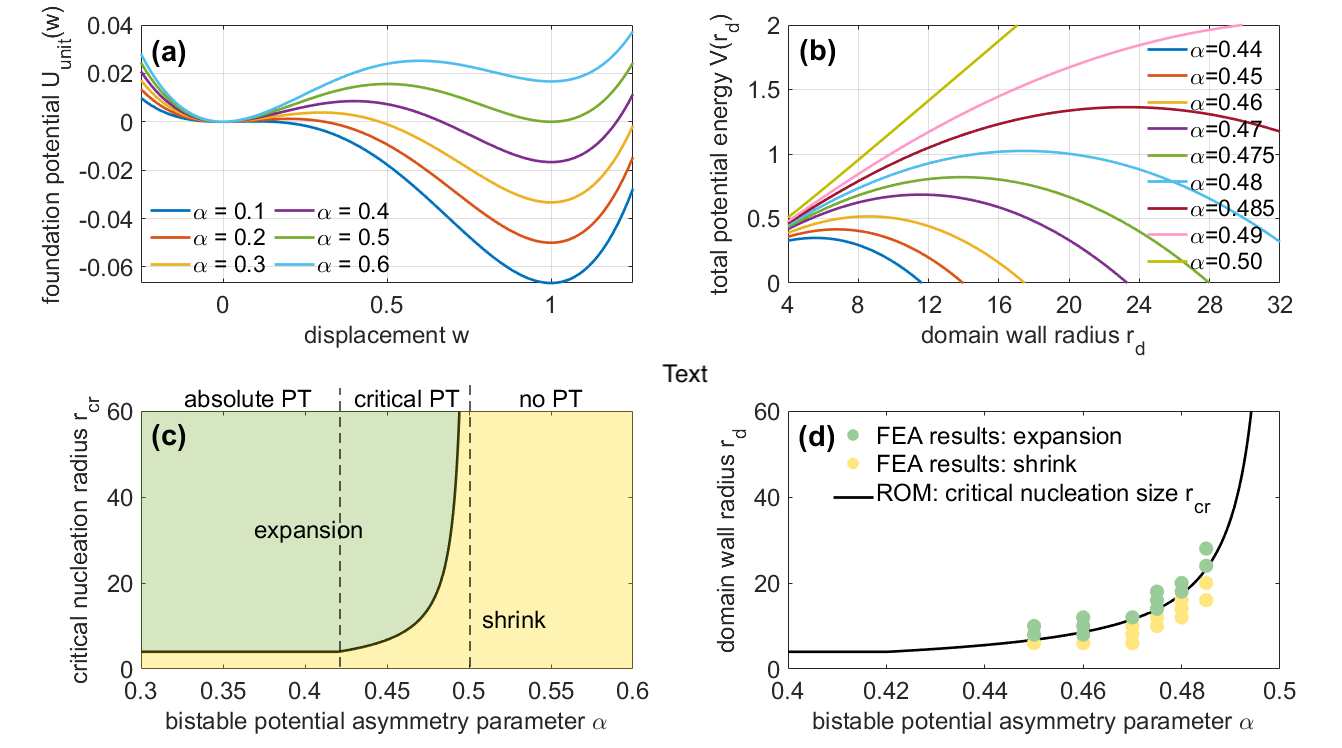}
  \caption{(a)~the potential energy landscape $U_{unit}(w)$ of a single bistable unit with different $\alpha$ values; (b)~the potential energy landscape $V(r_d)$ of the whole system by ROM with different $\alpha$ values; (c)~the critical nucleation size $r_{cr}$ as a function of $\alpha$ by the ROM Eq.~(\ref{r_cr}); (d)~comparison between the ROM and the FEA simulation results with different $\alpha$ and initial domain wall size $r_{\text{d0}}$. In FEA, $\gamma=\frac{1}{16}$.}
  \label{contiunous energy}
\end{figure}

The critical nucleation size is important as it can be used to determine the minimum size of the actuation area required to trigger a global state change in the system. The critical nucleation size $r_{cr}$ is the position of the global maximum in the potential energy landscape and thus can be determined by:
\begin{equation}
\frac{dV(\alpha, r_d)}{dr_d}|_{r_d=r_{cr}} = 0
\end{equation}
Using Eq.~(\ref{potential energy}), the critical nucleation size $r_{cr}$ is the largest root of a cubic equation:
\begin{equation}\label{r_cr}
\frac{2\alpha-1}{6}r_{cr}^3+\left( \frac{W}{96}+\frac{256}{15W^3} \right)r_{cr}^2-\frac{4}{3W}=0
\end{equation}  
In some cases, the largest root lies in Stage 2, or the equation yields only a negative (nonphysical) root. In either situation,  $\frac{\partial V(r_d)}{\partial r_d}<0$ holds for all $r>\frac{W}{2}=4$, implying that an initial domain wall with $r_d=\frac{W}{2}$ is sufficient to initiate a global phase transition. For these cases we set $r_{cr}=4$ for simplicity. The dependence of $r_{cr}$ on $\alpha$ is shown in Fig.~\ref{contiunous energy}. 

As \( \alpha \) increases, the critical nucleation size \( r_{cr} \) grows significantly. The system can be categorized into three distinct regimes according to the value of \( \alpha \): 
\begin{enumerate}
\item \textbf{Absolute phase transition region:} ($\alpha<\alpha_1=0.42$): an initial domain wall with $r_d=W/2$ is sufficient to initiate a global phase transition, and creating a domain wall $r_d=\frac{W}{2}$ only requires a point load. This threshold $\alpha_1$ can be determined analytically from Eq.~(\ref{r_cr}) when the largest root is equal to $\frac{W}{2}$.
\item \textbf{Critical phase transition region:} ($0.42=\alpha_1<\alpha<\alpha_2=0.5$): there exists a critical domain wall size $r_{cr}$ for which the domain wall expands when the initial domain wall size  $r_{d0}>r_{cr}$ and shrinks and finally disappears when the initial domain wall size $r_{d0}<r_{cr}$.
\item \textbf{No phase transition region:} ($\alpha>=0.5$): the domain wall always shrinks to extinction, regardless of its initial size.
\end{enumerate}

The ROM predictions for $r_{cr}$ are validated against FEA in Fig.~\ref{contiunous energy}(d). In FEA, the discreteness parameter is selected as $\gamma=\frac{1}{16}$. For each $\alpha$ value, simulations with various initial radii $r_{d0}$ are performed and large damping coefficients are used to eliminate the effect of initial kinetic energy when the domain wall is released. Green markers denote cases where the domain wall expands, while yellow markers correspond to shrinkage and disappearance. The close agreement between the ROM predictions and the FEA results confirms the accuracy of the reduced-order model in capturing the nucleation threshold.

\subsubsection{Equation of motion: dynamic domain wall motion}
In the previous section, we established the criterion distinguishing between shrinking and expanding domain walls. We now turn to the dynamics, specifically, the rate at which the domain wall moves. To study the dynamic evolution, we formulate the Lagrangian of the reduced-order model in terms of its kinetic and potential energy:
\begin{equation}\label{lagrangian}
{L(r_d, \dot{r}_d)} = {E}_k(r_d, \dot{r}_d) - {V}(r_d)    
\end{equation}
In the ROM stage 1, only \( r_d(t) \) evolves and thus the Lagrangian depends only on \( r_d \) and its time derivative \( \dot{r}_d \).  The equation of motion then follows from the Euler–Lagrange equation
\begin{equation}\label{euler-lag eq}
\frac{d}{dt}\left( \frac{\partial {L}}{\partial \dot{r}_d} \right) - \frac{\partial {L}}{\partial r_d} = 0    
\end{equation}
To obtain the Lagrangian $L$ explicitly, next we derive the kinetic energy $E_k$ . The velocity field of the plate when a domain wall is shrinking or expanding can be obtained from Eq.~(\ref{tanh1}):
\begin{equation}\label{velocity}
v(r,t) = \frac{\partial w}{\partial t} = \frac{2w_b}{W} \text{sech}^2\left( \frac{r - r_d(t)}{W/4} \right) \dot{r}_d
\end{equation}
and the corresponding kinetic energy is:
\begin{equation}\label{eq: kinetic main}
\begin{aligned}
E_{k1} &= \iint \frac{1}{2} \rho h v^2 \, dA \\
  &= \frac{4\pi \rho h w_b^2}{W^2} \dot{r}_d^2 \int_0^\infty \text{sech}^4\left( \frac{r - r_d}{W/4} \right) r \, dr \\
  &= \frac{4\pi \rho h w_b^2 }{3} \frac{r_d}{W} \dot{r}_d^2
\end{aligned}
\end{equation}
implying an effective mass,  
\begin{equation}
m_{\mathrm{eff}} = \frac{8\pi \rho h w_b^2 }{3} \frac{r_d}{W},
\end{equation}
which scales with the domain wall radius \(r_d\) because the kinetic energy is localized near the wall and proportional to its perimeter. The same integration techniques for sech-based functions used in deriving the plate and foundation energies in Eqs.~(\ref{eq: plate energy main 1}) and (\ref{eq:foundation energy main 1}) are also applied in Eq.~(\ref{eq: kinetic main}). The complete derivation is included in~\ref{append B}.

Substituting Eq.~(\ref{eq: kinetic main}) into the Euler--Lagrange equation (\ref{euler-lag eq}) yields  
\begin{equation}\label{ODE 1}
2r_d \ddot{r}_d
+ \dot{r}_d^2
- \frac{1}{r_d^2}
+ \frac{W}{8}(2\alpha - 1) r_d 
+ \frac{64}{5W^2}
+ \frac{W^2}{128}= 0 .
\end{equation}
A first integration of Eq.~(\ref{ODE 1}) gives  
\begin{equation}\label{v-r relation}
    \dot{r}_d^{\,2} = 
- \frac{W}{16}(2\alpha - 1)\, r_d
- \frac{1}{r_d^{2}} - \frac{64}{5 W^2} - \frac{W^2}{128}
+ \frac{C}{r_d} ,
\end{equation}
with the integration constant \(C\) determined by the initial condition \((r_{d0},\dot r_{d0})\) as  
\begin{equation}
C = r_{d0} \left[ 
 \frac{W}{16}(2\alpha - 1)\, r_{d0} + \frac{1}{r_{d0}^{2}}
+ \frac{64}{5 W^2} + \frac{W^2}{128} + \dot{r}_{d0}^{\,2} 
 \right].
\end{equation}

\begin{figure}[H]
  \centering
  \includegraphics[width=0.65\linewidth]{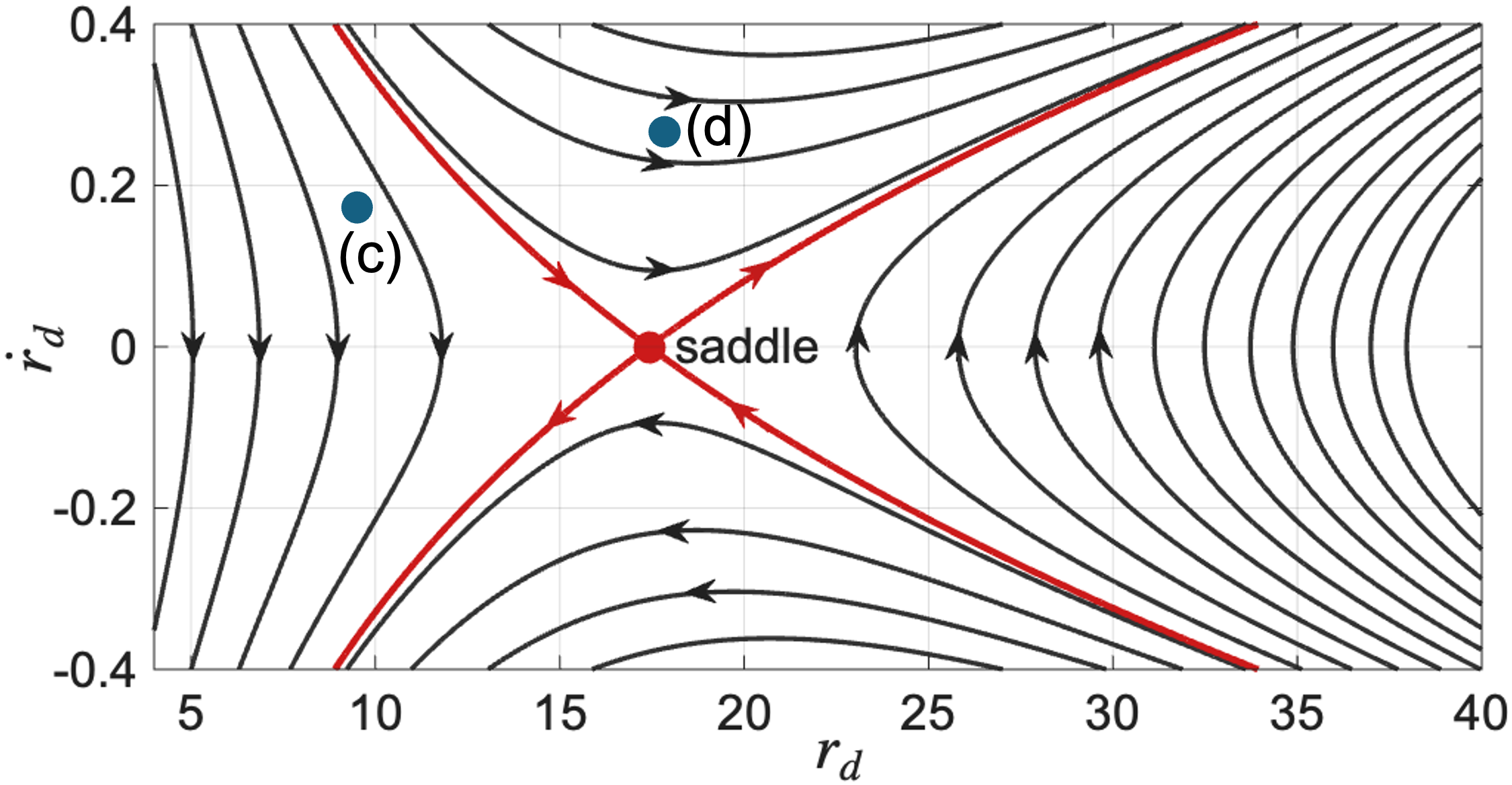}
  \caption{ROM trajectory of the dynamic evolution of a domain wall in the $(r_d,\dot r_d)$ space. The position of the saddle point corresponds to the critical nucleation size $r_{\text{cr}}$. Arrows denote the actual motion directions. Points (c) and (d) denote the initial conditions in fig.~\ref{dynamic_FEA_ROM} (c) and (d). With $\alpha=0.48$.}
  \label{v_r_relation}
\end{figure}

\begin{figure}[H]
  \centering
  \includegraphics[width=\linewidth]{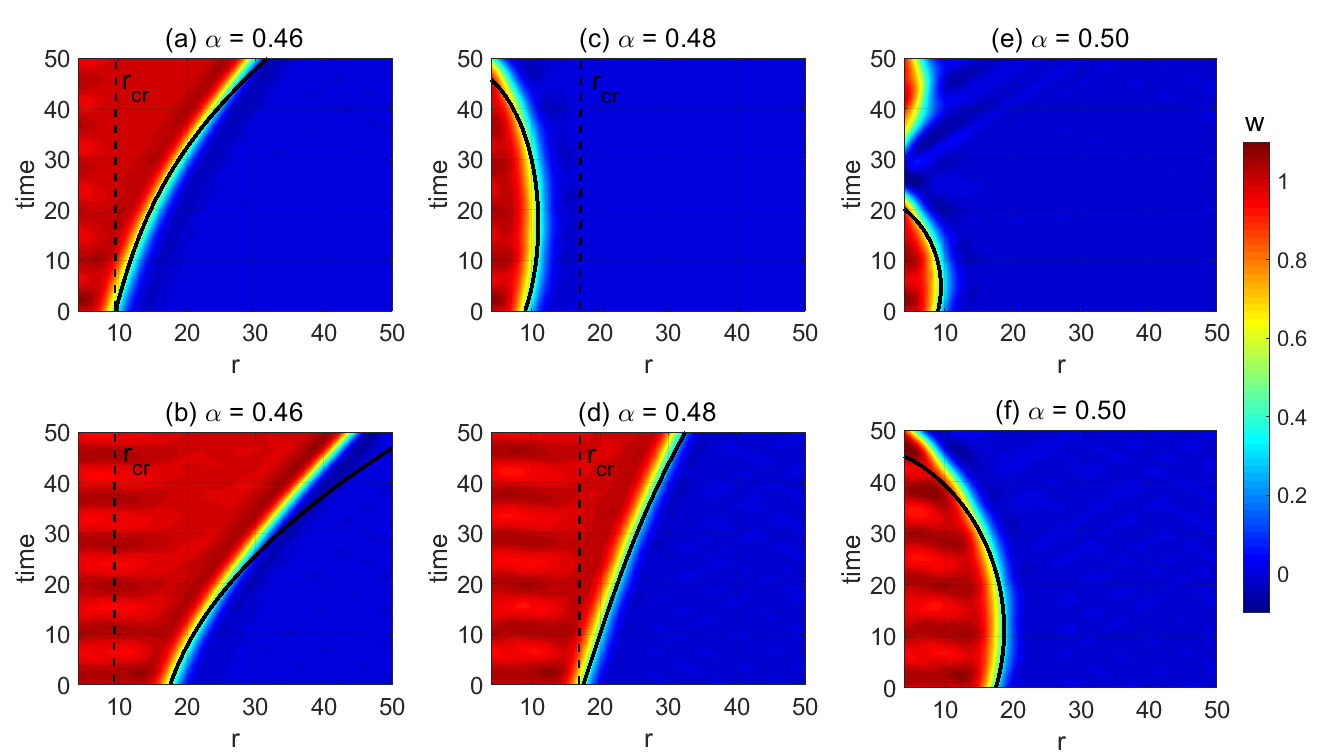}
  \caption{Domain wall dynamic evolution: displacement $w$ contours obtained from FEA simulations, with the black line showing the domain wall center position predicted by the ROM, Eq.~(\ref{ODE 1}). The dashed line denotes the critical nucleation size given by Eq.~(\ref{r_cr}).  In FEA, $\gamma=\frac{1}{16}$. }
  \label{dynamic_FEA_ROM}
\end{figure}

Eq.~(\ref{v-r relation}) is equivalent to the energy conservation, relating \(\dot r_d\) to \(r_d\) for any undamped motion. Fig.~\ref{v_r_relation} shows contours of constant \(C\) in the \((r_d, \dot r_d)\) phase space with parameter $\alpha=0.48$, each contour representing a trajectory of fixed total energy; arrows indicate the motion direction. The saddle point in Fig.~\ref{v_r_relation} corresponds to the critical nucleation size \(r_{\text{cr}}\), i.e., the global maximum of the potential energy landscape. For a wall released from rest, the trajectories show that \(r_{d0}<r_{\text{cr}}\) leads to shrinkage and \(r_{d0}>r_{\text{cr}}\) to expansion. The trajectories also show that if the domain wall has a sufficiently large initial velocity (\(\dot r_{d0}>0\)), it may overcome the potential barrier even when \(r_{d0}<r_{\text{cr}}\), leading to unbounded expansion. In all cases, the domain wall moving velocity magnitude \(|\dot r_d|\) increases as \(r_d\) moves away from \(r_{\text{cr}}\), regardless of whether the wall is shrinking or expanding.

To validate this dynamic ROM, we extracted the initial conditions $ r_{d}(0)$ and $\dot{r}_d(0)$ from FEA simulations and then integrated the ODE Eq.~(\ref{ODE 1}) using 4th-order Runge-Kutta method (\texttt{ode45} in Matlab) with these initial conditions. Fig.~\ref{dynamic_FEA_ROM} compares the ROM predictions with the FEA results, demonstrating close agreement for both shrinking and expanding cases.

\subsection{Transition from continuum limit to discrete case} \label{subsection 4.2}

\subsubsection{ROM prediction}
\begin{figure}[H]
  \centering
  \includegraphics[width=\linewidth]{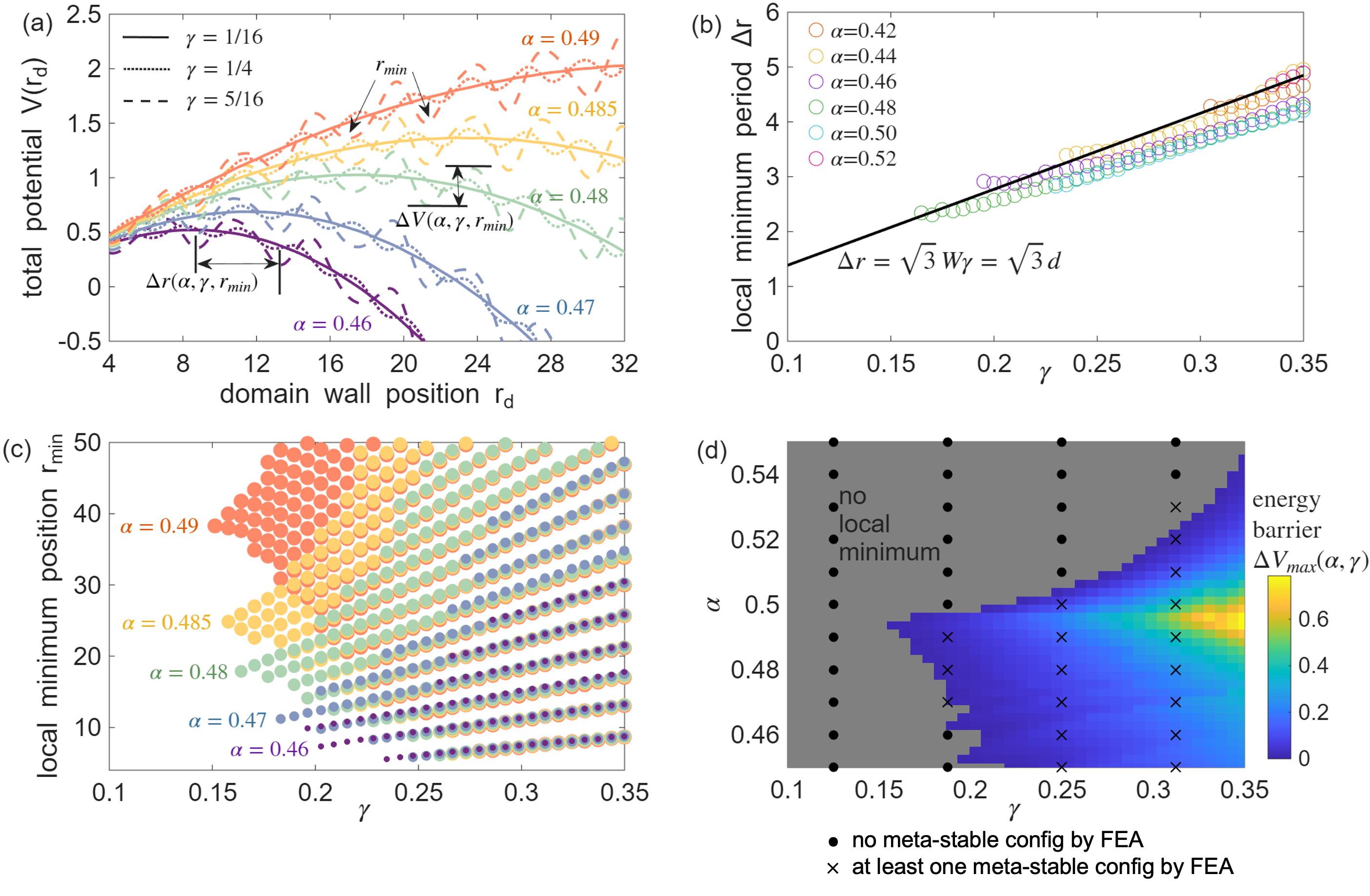}
  \caption{the ROM prediction for the effect of the discreteness $\gamma$ on the potential energy landscape and domain wall stability: (a)~the energy landscape $V(r_d)$; (b)~the spatial period $\Delta r$ of local minima in energy landscape is proportional to the foundation unit spacing $\gamma$; (c)~the spatial distribution of local minima in energy landscape; (d)~the maximum energy barrier $\Delta V_{max}$ in the potential energy landscape.}
  \label{discrete energy}
\end{figure}

Fig.~\ref{discrete energy} presents the reduced-order model (ROM) prediction for the effect of the discreteness $\gamma$ on the potential energy landscape and domain wall stability. In Fig.~\ref{discrete energy}(a), the total potential energy  \( V(r_d) \) is plotted for a range of bistable potential parameters \( \alpha = 0.46, 0.47, 0.48, 0.485, 0.49 \) and discreteness parameters \( \gamma = 0.1, 0.25, 0.35 \). At scales much larger than the lattice size \( d \), the energy landscape is primarily governed by the competition between the bulk energy release and surface energy cost. However, at finer scales (on the order of $d$), the underlying lattice discreteness $\gamma$ gives rise to modulations in the potential energy.  For small \( \gamma \) (e.g., \( \gamma = 0.1 \)), the energy landscape is nearly identical to the continuous counterpart in Fig.~\ref{contiunous energy}. As \( \gamma \) increases, periodic oscillations emerge: the domain wall experiences a periodic variation in total energy as it moves relative to the discrete lattice (Fig.~\ref{P-N effect}), analogous to the Peierls–Nabarro (P-N) effect.

The dependence of spatial period \( \Delta r \) of local minima in Fig.~\ref{discrete energy}(a) on parameters $\alpha$ and $\gamma$ are plotted in Fig.~\ref{discrete energy}(b).  The spatial period \( \Delta r \) is largely independent of \( \alpha \) and primarily determined by \( \gamma \). The hexagon foundation lattice has a geometric periodicity of $\sqrt{3}d$ in the radial direction, marked by the black line in Fig.~\ref{discrete energy}(b). The close agreement between the observed energy landscape periodicity \( \Delta r \) and the lattice geometric periodicity confirms that the periodic modulation in the energy landscape is a direct consequence of the underlying discreteness of the foundation.

A domain wall near a local energy minimum $r_{\text{min}}$ experiences restoring forces from adjacent energy barriers (local maxima). As a result, small perturbations lead to bounded motion around $r_{\text{min}}$, and dissipation eventually traps the domain wall at such a metastable position. The degree of stability of such a state can be characterized by the energy barrier $\Delta V(\alpha, \gamma, r_{\text{min}})$, defined as the smaller of the two energy differences between a local minimum and its adjacent maxima.

Fig.~\ref{discrete energy}(c) and Fig.~\ref{discrete energy}(d) show the local minimum positions \( r_{\min} \) and the maximum energy barriers \( \Delta V_{\max} \) across various combinations of $\alpha$ and $\gamma$, respectively. For each $\alpha$, there exists a critical discreteness $\gamma_{\text{cr}}(\alpha)$, beyond which the P–N modulation produces at least one local minimum. Increasing $\gamma$ beyond $\gamma_{\text{cr}}$ generally leads to more minima and larger energy barriers, indicating stronger domain wall stability. Additionally, three key observations can be made from Figs.~\ref{discrete energy}(c) and Fig.~\ref{discrete energy}(d):

1. For each $\alpha$, the location of the local-minimum onset consistently aligns with $r_{\text{cr}}(\alpha)$, the location of the global maximum in the corresponding continuum energy curve (see Fig.~\ref{contiunous energy}(c) and Fig.~\ref{discrete energy}(d)). This alignment occurs because the energy curve is flattest near $r_{\text{cr}}(\alpha)$, making it most sensitive to the P-N modulation. This phenomenon can be more directly observed in Fig.~\ref{discrete: FEA_ROM}(a). 

2. For a set of $\gamma$ and $\alpha$ values, metastable states occur only within a finite range of $r_{\text{dst}}(\gamma, \alpha)$ (Fig.~\ref{discrete energy}(c)). As previously analyzed in Section~2, the periodic P-N energy modulation originates from the bistable units within the domain wall (Fig.~\ref{P-N effect}(b)). Therefore, the magnitude of the P-N modulation scales with $r$ (the perimeter of the domain wall). When $r$ is too small, the oscillation magnitude is insufficient to generate energy barriers, eliminating local minima; when $r$ is too large, the modulation (which scales with ~$r$) is negligible compared to the background energy at a larger scale (which scales as ~$r^2$), again suppressing metastability. Hence, metastable domain wall positions are confined to an intermediate range of $r_{\text{dst}}$.

3. In Fig.~\ref{discrete energy}(c) and Fig.~\ref{discrete energy}(d), the smallest value of \( \gamma_{\text{cr}}(\alpha)\) occurs as \( \alpha \to 0.5^- \). In the regime close to \( \alpha \to 0.5^- \) the continuous counterpart of the energy curve exhibits a wide flat region near its maximum, making the system particularly sensitive to the P-N effect. As \( \alpha \) deviates further from 0.5, a larger discreteness is required to induce the same degree of metastability.

\subsubsection{Comparison between ROM and FEA}
\begin{figure}[H]
  \centering
  \includegraphics[width=\linewidth]{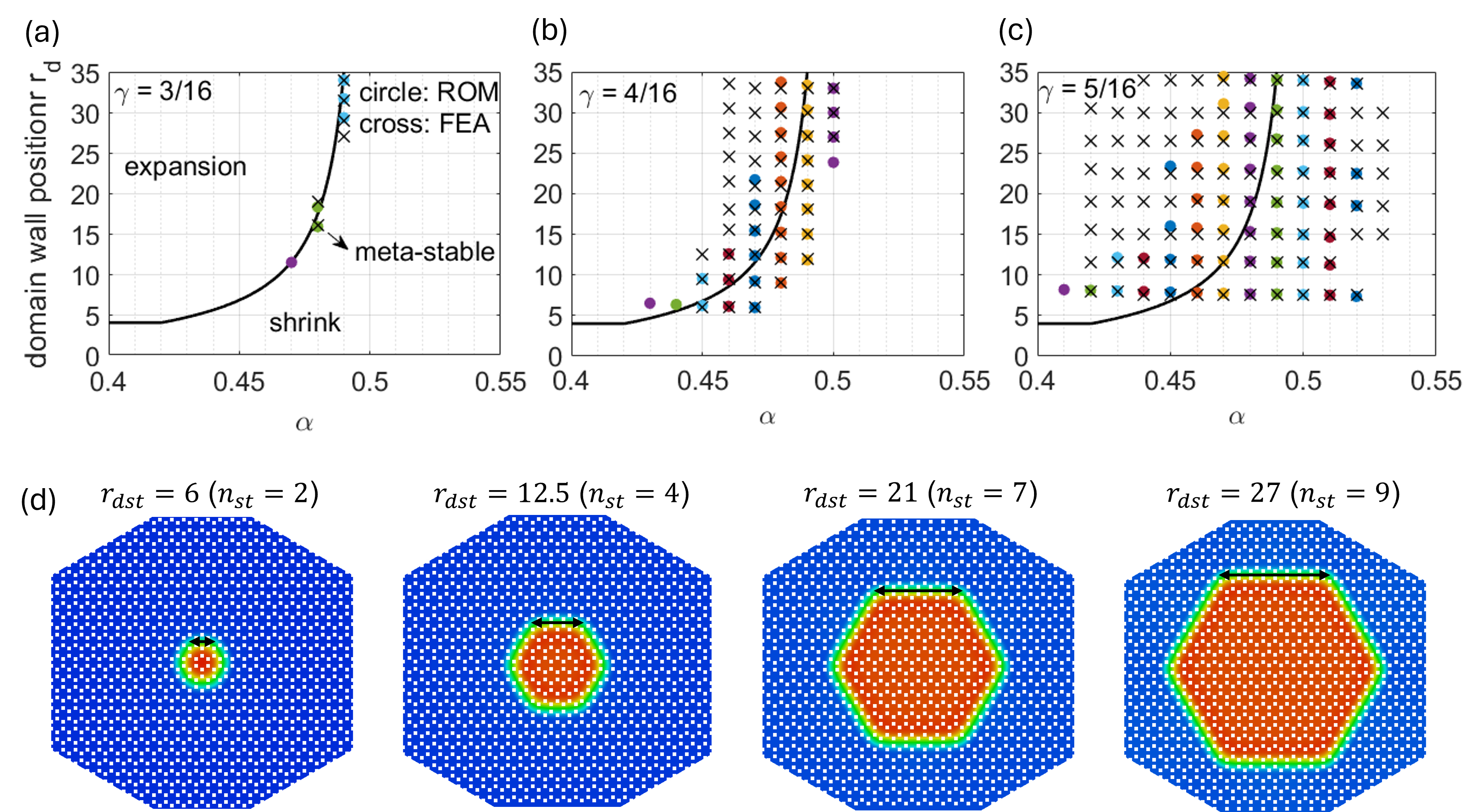}
  \caption{(a-c)~the stable domain wall positions predicted by ROM and FEA simulations for $\gamma=3/16, 4/16$, and $5/16$, respectively. (d)~the stable domain wall configurations in FEA with $\alpha=0.47$ and $\gamma=4/16$. The color contours in (d) represent the out-of-plane displacement, and the color bar is identical to that used in Fig.~\ref{fig1}}
  \label{discrete: FEA_ROM}
\end{figure}

To validate the ROM in the discrete case, we conducted a series of finite element simulations with different discreteness parameters, \( \gamma \). The simulation procedure was the same as described previously in Section 2. To identify all possible stable states, we simulated a range of initial radii \( r_{d0} \in (0, 35) \) for each $\gamma$ value.

For small values of $\gamma$, such as \( \gamma = \frac{1}{16} \) and \( \frac{2}{16} \), both FEA and ROM show that no metastable domain wall exists for any value of \( \alpha \). The behavior in this regime follows the continuous case, where the domain wall either shrinks or expands according to $r_{d0}$. Therefore, we focus on the cases with \( \gamma > \frac{2}{16} \). Fig.~\ref{discrete: FEA_ROM}(a)–Fig.~\ref{discrete: FEA_ROM}(c) compare FEA results with ROM predictions for \( \gamma = \frac{3}{16}, \frac{4}{16}, \frac{5}{16} \), where circles denote local minima position by ROM predictions, crosses represent stable domain wall radius in FEA, and the black line shows the critical radius \( r_{\text{cr}}(\alpha) \) from the continuous model (see Section~\ref{section 4.1.1} and Fig.~\ref{contiunous energy}). In Fig.~\ref{discrete: FEA_ROM}(a), FEA shows that the onset of stable configurations does occur near \( r_{\text{cr}}(\alpha) \), consistent with the ROM prediction. Both FEA and ROM show that for each \( \gamma \), there exists a range of \( \alpha \) values for which the system supports at least one stable domain wall radius. In Fig.~\ref{discrete energy}(d) and Fig.\ref{discrete: FEA_ROM}(a)-(c), this stable range of $\alpha$ broadens with increasing \( \gamma \), and the ROM predictions agree well with the FEA results. Besides, both FEA and ROM show that the stable range of domain wall radius \( r_{\text{dst}} \) also broadens with larger \( \gamma \). The ROM predicts the lower bound of  \( r_{\text{dst}} \) quite well but underestimates its upper bound, indicating that ROM tends to underestimate the domain wall stability for large radii.

The discrepancy between FEA and ROM for large \( r_d \) arises from the ROM’s assumption of a perfectly circular domain wall. As shown in Fig.~\ref{discrete: FEA_ROM}(d), FEA results show that small domain walls is more similar to a circular shape, aligning well with ROM assumption. However, as \( r_d \) increases, the final shape of the domain wall increasingly conforms to the geometry of the hexagonal foundation lattice. For sufficiently large radii, the system evolves into regular hexagonal domain walls, which minimize energy by optimizing alignment with the foundation lattice, giving lower energy than a circular wall of approximately the same size. Consequently, even though the ROM predicts instability for large circular domain walls, the actual system stabilizes by adopting a hexagonal geometry, leading to underestimation of the upper bound of the stable \( r_{\text{dst}} \) range in the ROM.

The FEA results in Fig.~\ref{discrete: FEA_ROM}(d) also reveal that the stable hexagonal domain walls have side lengths
\[
D = \sqrt{3} d n_{\text{st}}
\]
where \( \sqrt{3}d \) is the spacing between centers of neighboring hexagonal lattices and \( n_{\text{st}} \) is an integer. Moreover, the orientation of these stable hexagons is aligned with the orientation of the foundation lattice. Based on the stable range of \( r_{\text{dst}} (\gamma, \alpha) \), one can obtain the corresponding range of stable integers \( n_{\text{st}}(\gamma, \alpha) \). Describing the stable configurations in terms of \( n_{\text{st}} \) rather than \( r_{\text{dst}} \) offers a more convenient geometric parameterization.

\begin{figure}[H]
  \centering
  \includegraphics[width=\linewidth]{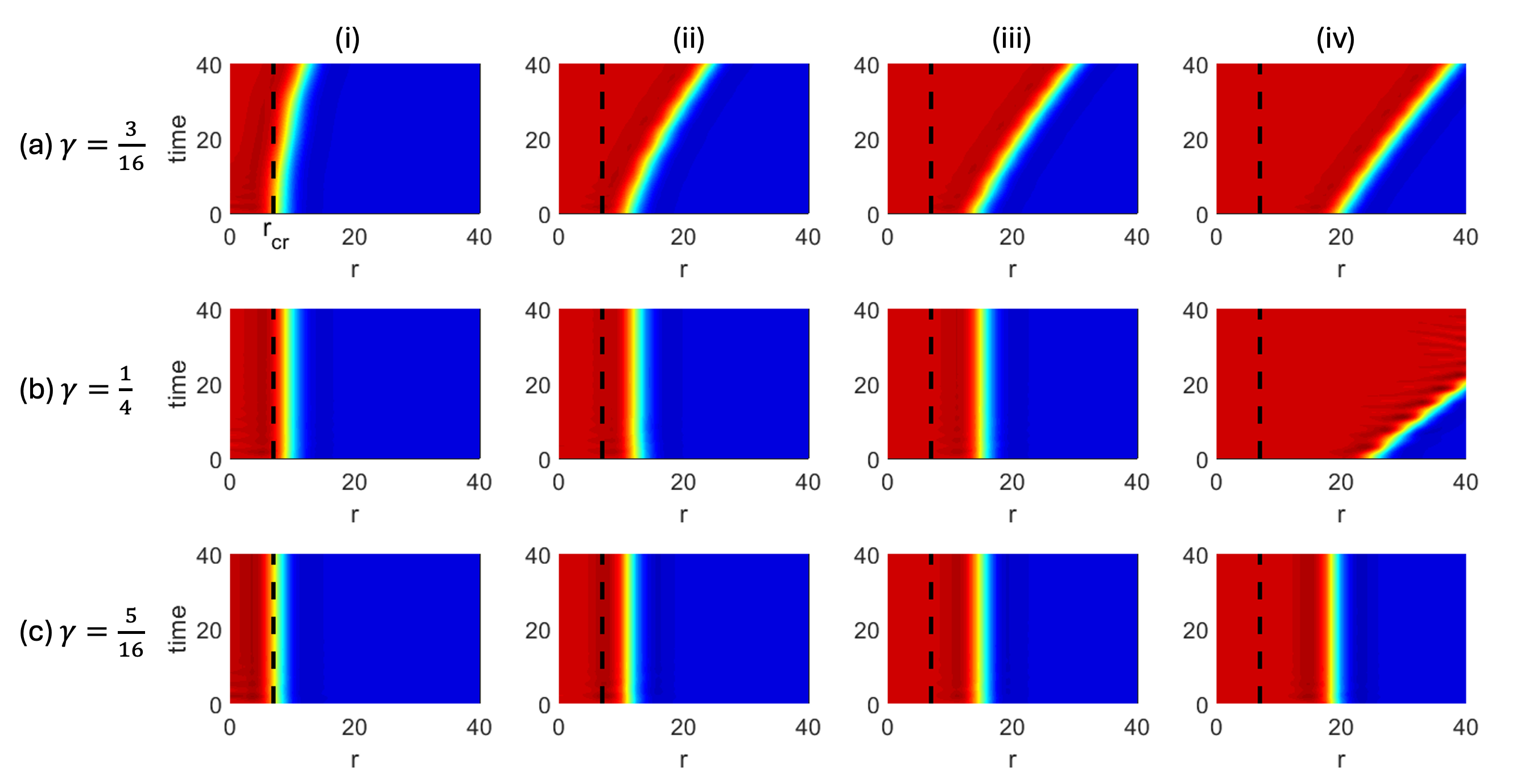}
  \caption{the domain wall evolution history in FEA for different discreteness values $\gamma$. With asymmetry parameter~$\alpha=0.45$. The dashed line represents $r_{cr}$. The color represent the out-of-plane displacement  $w$, and the color bar is identical to that used in Fig.~\ref{fig1}.}
  \label{discrete3}
\end{figure}

\begin{figure}[H]
  \centering
  \includegraphics[width=\linewidth]{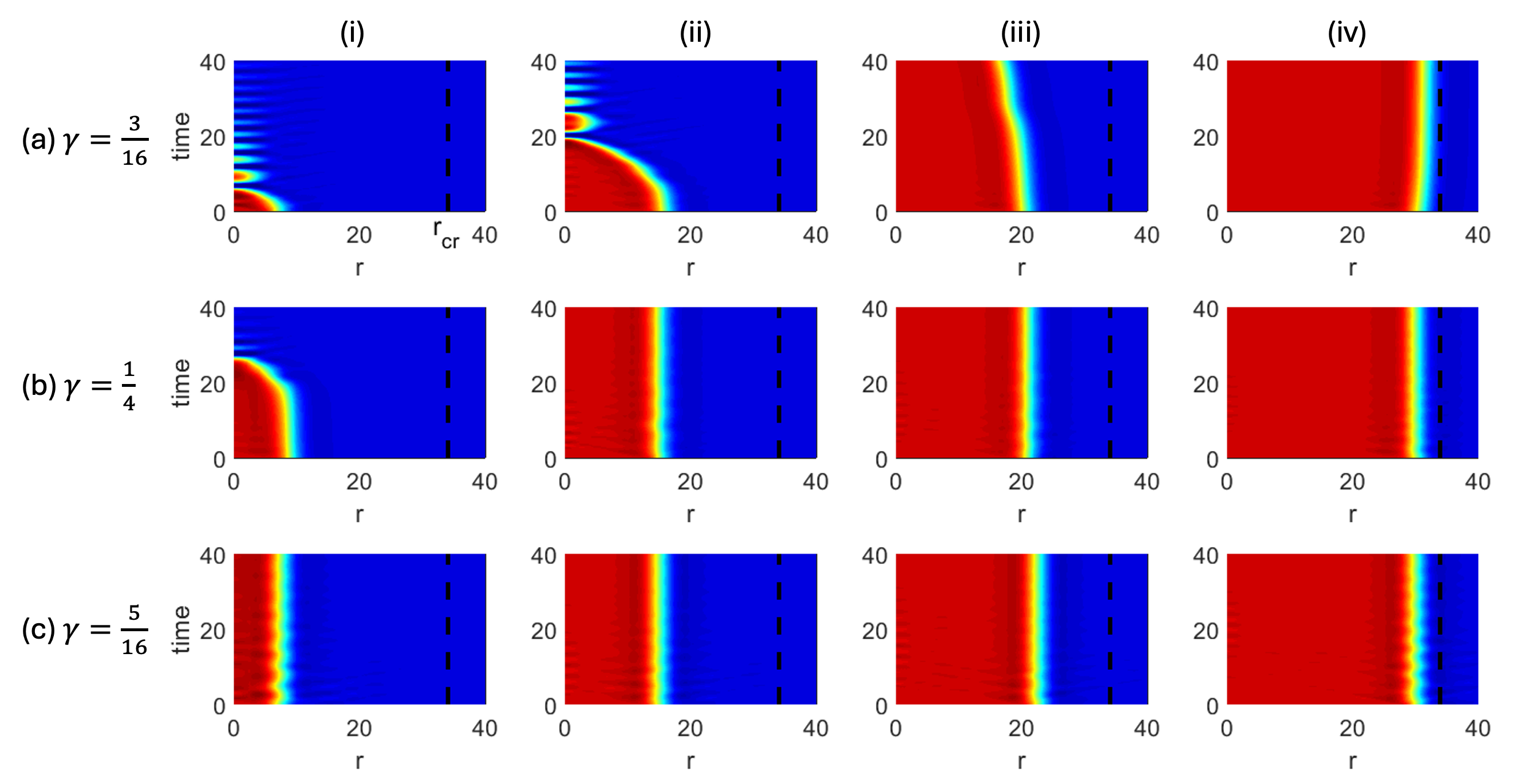}
  \caption{the domain wall evolution history in FEA for different discreteness values $\gamma$. With asymmetry parameter~$\alpha=0.49$. The dashed line represents $r_{cr}$.  The color represent the out-of-plane displacement $w$, and the color bar is identical to that used in Fig.~\ref{fig1}.}
  \label{discrete4}
\end{figure}

Outside the stable \( r_{\text{dst}} \) range, the system exhibits behaviors analogous to those in the continuous case. Specifically, if the initial size \( r_{d0} \) is smaller than the lower bound of \( r_{\text{dst}} \), the domain wall shrinks and vanishes. Conversely, if \( r_{d0} \) exceeds the upper bound of \( r_{\text{dst}} \), the wall expands and fails to stabilize. Fig.~\ref{discrete3} and \ref{discrete4} illustrate representative behaviors: shrinking (Fig.~\ref{discrete4}(ai)-(aiii) and (bi)), stabilization to a metastable state (Fig.~\ref{discrete3}(bi)-(biii) and (c), Fig.~\ref{discrete4}(aiv)(bii)-(biv) and (c)), and expansion (Fig.~\ref{discrete3}(a) and (biv)).

\section{General polygon domain walls}\label{section 5}

In the continuous limit, for a propagating domain wall 
the circular shape is the most energy-efficient configuration for front evolution, 
minimizing the increase in perimeter and thereby the surface energy cost that is required for an increase in area. 
When the initial nucleation shape deviates from circularity, such as being elliptical or irregular, the system naturally evolves toward a more circular configuration. As discussed in the previous section, a certain degree of foundation discreteness helps pin the axisymmetric (or regular hexagonal) domain wall, preventing it from moving, which makes it possible to have a stable irregular domain wall shape in this system. In this section, we explore the stability of domain walls in irregular shapes and connect the stability of irregular shapes with the axisymmetric (regular hexagonal) case. 

\subsection{Irregular convex hexagon with internal angles of \( 120^\circ \)}
As shown in Section~\ref{section 4}, for given values of \( \alpha \) and \( \gamma \), there exists a range of integers \( n_{\text{st}} \) within which regular hexagonal domain walls with side length \( D = \sqrt{3}d n_{\text{st}}\) are stable. As previously discussed, the prominence of the P–N effect scales with the domain wall perimeter and thus with the side length \( D \). Only within this stable range of \( n_{\text{st}} \) does the P–N effect sufficiently modulate the energy landscape to enable metastable configurations. 
Due to the six-fold symmetry of the regular hexagon, the global stability of the entire domain wall is equivalent to the local stability of a segment defined by a single edge of length \( D = \sqrt{3}d n_{\text{st}} \) and two adjoining interior angles of \( 120^\circ \). This local stability criterion can be extended to assess the global stability of irregular hexagonal domain walls whose interior angles are \( 120^\circ \) but side lengths differ. 

We define a general irregular hexagon with all the interior angles of \( 120^\circ \) and side lengths of \( D_i = \sqrt{3}d n_i\), where \( n_i \) are integers and $i=1, ..., 6$. 
We expect that if all \( n_i \) fall within the stable range of \( n_{\text{st}} \) for given values of \( \alpha \) and \( \gamma \), then all edges individually satisfy the P–N stability condition, and thus the entire irregular hexagon domain wall remains stable. Conversely, if any of the side lengths fall outside the stable range, the corresponding edges will be energetically unfavorable. Those exceeding the upper bound will tend to expand (move outwards), while those below the lower bound will contract (move inwards). Those edges ultimately lead to the distortion, expansion, or collapse of the irregular hexagonal domain wall.

\begin{figure}[H]
  \centering
  \includegraphics[width=\linewidth]{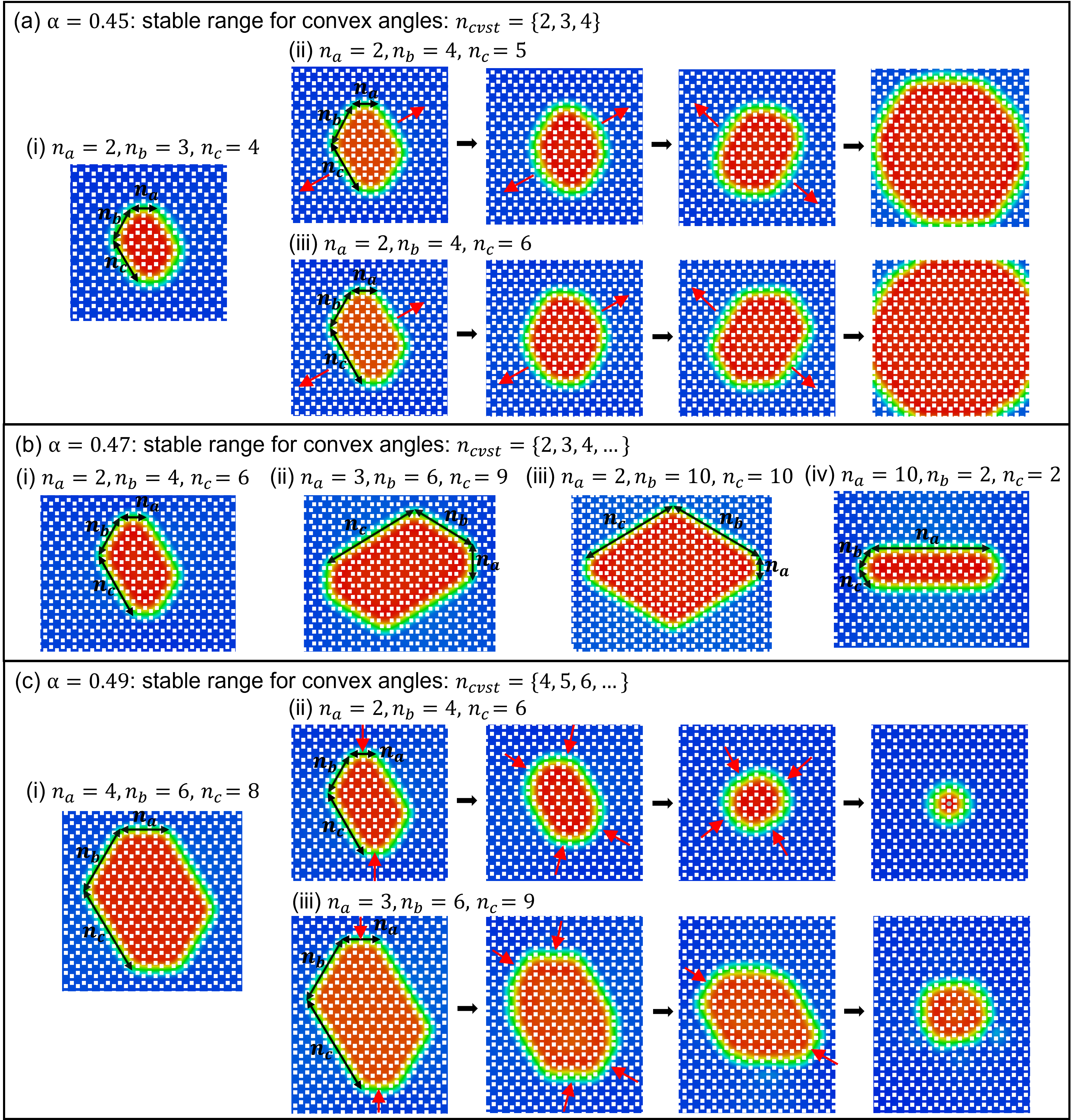}
  \caption{FEA results for irregular hexagonal domain wall evolution at $\gamma = \frac{1}{4}$:  
(a)~$\alpha = 0.45$: (i) stable configuration; (ii), (iii) unstable and expanding cases;  
(b)~$\alpha = 0.47$: (i)–(iv) all stable configurations;  
(c)~$\alpha = 0.49$: (i) stable; (ii), (iii) unstable and shrinking cases. The color represent the out-of-plane displacement $w$, and the color bar is identical to that used in Fig.~\ref{fig1}.
}
  \label{irhex1}
\end{figure}

The above reasoning assumes that the edges of the irregular hexagon behave independently in terms of their stability, and thus the stability of irregular configurations can be inferred from the stability of regular hexagons. This simplification enables a convenient extension of the results obtained in the regular hexagonal cases. However, in reality, adjacent edges and interior angles are geometrically and elastically coupled. The mechanical response of one edge can influence the stresses and deformations in neighboring edges through these geometric constraints. As a result, 
the stability inference presented above should be regarded as a first-order approximation. To assess its validity, we performed FEA simulations on irregular hexagons with side lengths selected near the boundaries of the stability range obtained from regular hexagons. We generalized the irregular hexagon geometry by introducing three distinct edge lengths, while preserving the internal angle of \( 120^\circ \) at each vertex. The domain wall was constructed as a closed hexagon with the following structure: two edges of length \( A = \sqrt{3}d n_a \), two of \( B = \sqrt{3}d n_b  \), and two of \( C = \sqrt{3}d n_c \), where \( n_a, n_b, n_c \in \mathbb{Z}^+ \) were integer multipliers, and \( d \) was the edge length of the foundation unit cell. We selected $\gamma=1/4$ as with this $\gamma$ value the lower bounds and upper bounds of  $n_{\text{st}} $ for different $\alpha$ were distinguished and well separated. The representative cases in our FEA simulation are shown in Table~\ref{tab:general_irregular_cases}. The stability of domain walls in FEA simulations are consistent with the inferred stability, as shown in Fig.~\ref{irhex1}.

\begin{table}[h!]
\centering
\caption{Summary of FEA cases for irregular hexagonal domain wall with convex internal angles \(120^\circ\). The edge lengths of the hexagon are \(D = \sqrt{3}d n\). Discreteness parameter is $\gamma=\frac{1}{4}$.}
\begin{tabular}{|c|c|c|c|c|l|}
\hline
\(\alpha\) & stable range \(n_{\text{cvst}}\) for \(120^\circ\) angles & \(n_a\) & \(n_b\) & \(n_c\) & \textbf{inferred stability} \\
\hline
\multirow{3}{*}{0.45} & \multirow{3}{*}{\(\{2, 3, 4\}\)} & 2 & 3 & 4 & stable\\
                      &                                 & 2 & 4 & 5 & edge \(C\) expands\\
                      &                                 & 2 & 4 & 6 & edge \(C\) expands\\
\hline
\multirow{4}{*}{0.47} & \multirow{4}{*}{\(\{2, 3, 4, \dots\}\)} & 2 & 4 & 6 & stable\\
                      &                                        & 3 & 6 & 9 & stable\\
                      &                                        & 2 & 10 & 10 & stable\\
                      &                                        & 10 & 2 & 2 & stable\\
\hline
\multirow{3}{*}{0.49} & \multirow{3}{*}{\(\{4, 5, 6, \dots\}\)} & 4 & 6 & 8 & stable\\
                      &                                        & 2 & 4 & 6 & edge \(A\) shrinks\\
                      &                                        & 3 & 6 & 9 & edge \(A\) shrinks\\
\hline
\end{tabular}
\label{tab:general_irregular_cases}
\end{table}

\subsection{Irregular polygon with concave internal angles of \( 240^\circ \)}
\begin{figure}[H]
  \centering
  \includegraphics[width=0.9\linewidth]{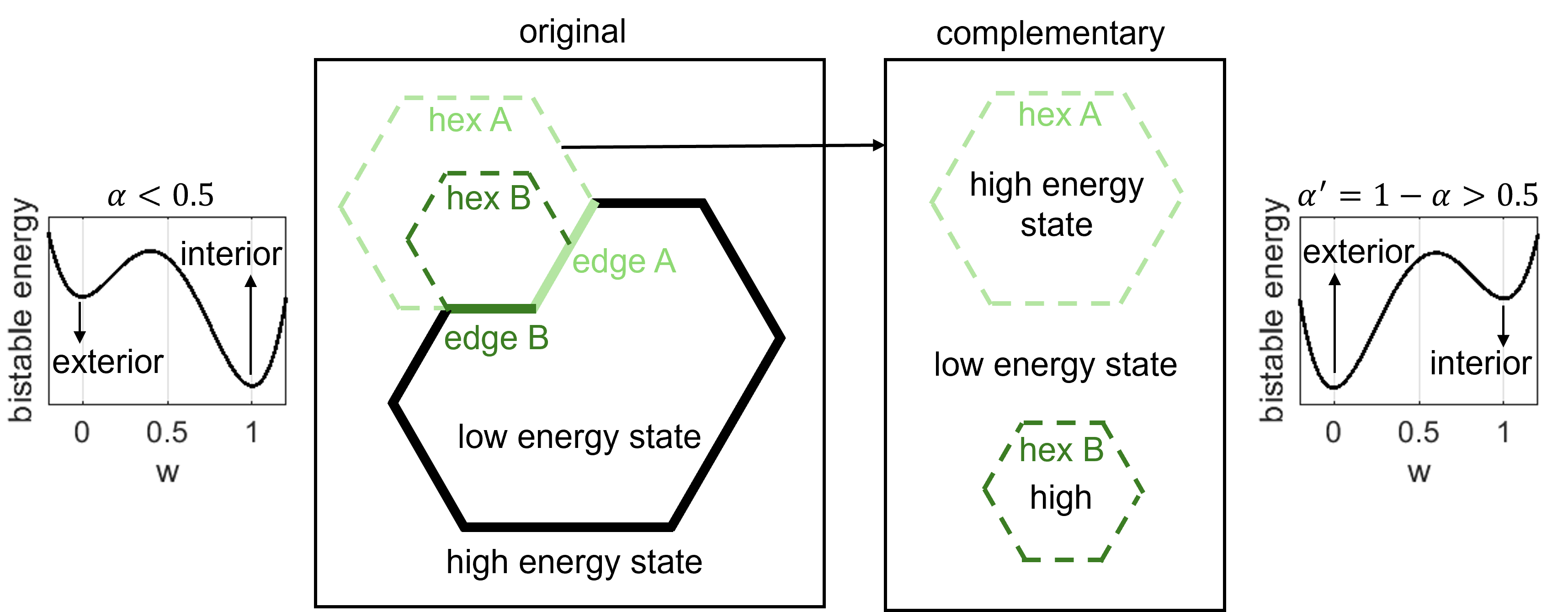}
  \caption{A domain wall in the shape of an irregular octagon with convex internal angles of \(120^\circ\) and one concave internal angle of \(240^\circ\), in a system with bistable foundation parameter \(\alpha\). The stability of the concave angle can be inferred from the stability of two complementary regular hexagons A and B with parameter $\alpha'=1-\alpha$ that conform to the concave corner.
}
  \label{irpoly1}
\end{figure}

\begin{figure}[H]
  \centering
  \includegraphics[width=1.0\linewidth]{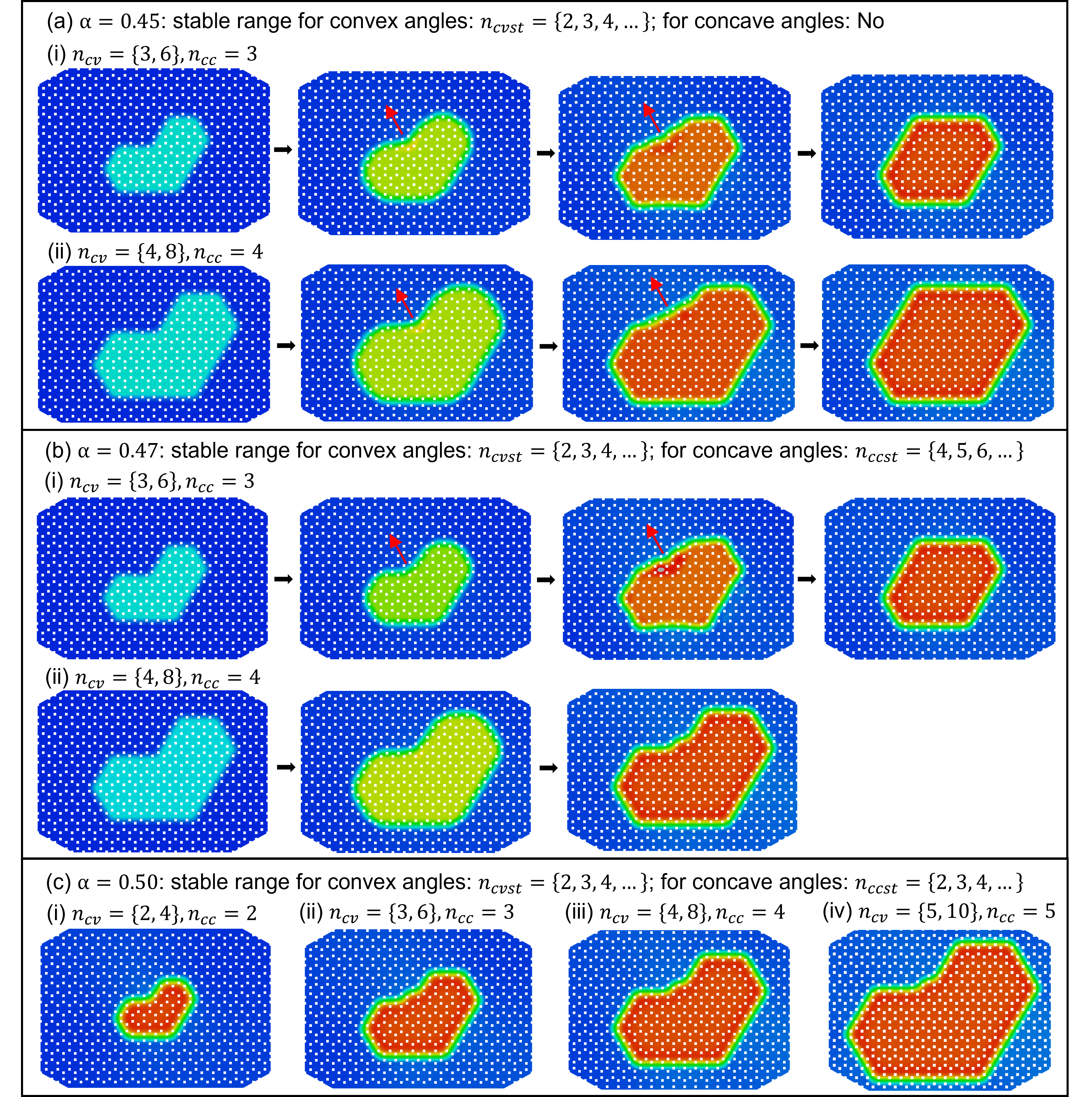}
  \caption{FEA results of irregular octagonal domain wall evolution at \(\gamma = \frac{5}{16}\):  
(a)~\(\alpha = 0.45\): (i)(ii) concave angles are unstable and evolve into convex angles;  
(b)~\(\alpha = 0.47\): (i) concave angle is unstable and evolves into a convex angle; (ii) stable concave angle;  
(c)~\(\alpha = 0.50\): (i)–(iv) all configurations exhibit stable concave angles. The color represent the out-of-plane displacement $w$, and the color bar is identical to that used in Fig.~\ref{fig1}.
}
  \label{irpoly2}
\end{figure}

To explore a broader class of potentially stable domain wall geometries, we now consider 
polygons whose internal angles are either \( 120^\circ \) or \( 240^\circ \), which are both geometrically compatible with the underlying hexagonal foundation lattice. For edges adjacent to concave angles, we propose the following stability inference. Suppose the foundation inside the polygon is in the low-energy state (i.e., \( \alpha < 0.5 \)), as shown in Fig.~\ref{irpoly1}. For a concave corner (such as the one formed by edges A and B), we construct two conceptual regular hexagons using the same edge lengths as A and B. To maintain the same relative energy distribution on both sides of edge A and B, the interior of each conceptual hexagon must correspond to the high-energy state, while the exterior corresponds to the low-energy state. This inverted energy distribution implies that the conceptual hexagons are governed by a complementary bistable parameter $\alpha'=1-\alpha$, leveraging the symmetry of the bistable potential Eq.~(\ref{bistable potential}) about $\alpha=0.5$. We denote these two conceptual regular hexagons as the complementary hexagons of the concave internal angle. If both of the complementary hexagons are stable under the complementary bistable parameter \( \alpha' \), we infer that the original edge pair forming the \( 240^\circ \) concave angle is also stable. This inference extends the same local-to-global reasoning previously applied in the convex case and relies on the assumption that edge stability is primarily determined by local bistable interactions and geometric compatibility. 
Now we apply the above reasoning to our system. From Fig.~\ref{discrete: FEA_ROM}(b), we know that for \( \gamma = \frac{1}{4} \), the stable range of \( \alpha \) for regular hexagons and therefore for convex \( 120^\circ \) internal angles is approximately \( [0.45, 0.5] \). The corresponding complementary bistable parameters are \( \alpha' = 1 - \alpha \in [0.5, 0.55] \). According to our previous inference, only the value \( \alpha' = 0.5 \) lies within the known stability range for regular hexagons. Therefore, we predict that only a system with \( \alpha = 0.5 \) can support stable polygons with both convex and concave angles with this value of \( \gamma \). For all other values \( \alpha' > 0.5 \), the corresponding complementary hexagons tend to shrink inward, since the critical domain wall size \( r_{\text{cr}} \to \infty \) for \( \alpha > 0.5 \). As a result, the original edges adjacent to concave angles tend to expand outward, making the configuration unstable. For a larger discreteness value \( \gamma = \frac{5}{16} \), Fig.~\ref{discrete: FEA_ROM}(c) shows that the stable range for convex \( 120^\circ \) internal angles broadens to \( [0.42, 0.53] \). In this case, both the original parameter \( \alpha \) and its complement \( \alpha' = 1 - \alpha \) must simultaneously lie within this range \( [0.42, 0.53] \) to ensure stability of both convex and concave angles. This leads to the necessary condition for a stable polygon with both convex and concave angles: \( 0.47 \leq \alpha \leq 0.53\). The full set of sufficient conditions, however, also depends on the edge lengths, as the stability of each segment is determined by its length. To validate this stability inference, we considered octagons with one concave angle of \(240^\circ\) and seven convex angles \(120^\circ\) and conducted FEA simulations. The representative cases are shown in Table~\ref{tab:fea_octagon_cases}, along with the predicted stable range of edge lengths for convex and concave angles.  The stability of domain walls in FEA simulation are consistent with the inferred stability, as shown in Fig.~\ref{irpoly2}.

\begin{table}[h!]
\centering
\caption{Summary of FEA cases for irregular octagonal domain wall with convex angles of \(120^\circ\) and one concave angle of \(240^\circ\). The edge lengths of the hexagon are \(D = \sqrt{3}d n\). Discreteness parameter is $\gamma=\frac{5}{16}$.}
\begin{tabular}{|c|c|c|c|c|l|}
\hline
\(\alpha\) 
& \makecell{stable range \(n_{\text{cvst}}\)\\for \(120^\circ\) convex angles} 
& \makecell{stable range \(n_{\text{ccst}}\)\\for \(240^\circ\) concave angles} 
& \(n_{\text{cv}}\) 
& \(n_{\text{cc}}\) 
& inferred stability \\
\hline

\multirow{2}{*}{0.45}
& \multirow{2}{*}{\(\{2, 3, 4, \ldots\}\)} 
& \multirow{2}{*}{No}
& \{3, 6\} & 3 & concave angle is unstable \\
& & & \{4, 8\} & 4 & concave angle is unstable \\
\hline

\multirow{2}{*}{0.47}
& \multirow{2}{*}{\(\{2, 3, 4, \ldots\}\)} 
& \multirow{2}{*}{\(\{4, 5, \ldots, 8\}\)}
& \{3, 6\} & 3 & concave angle is unstable \\
& & & \{4, 8\} & 4 & stable \\
\hline

\multirow{4}{*}{0.50}
& \multirow{4}{*}{\(\{2, 3, 4, \ldots\}\)} 
& \multirow{4}{*}{\(\{2, 3, 4, \ldots\}\)}
& \{2, 4\} & 2 & stable \\
& & & \{3, 6\} & 3 & stable \\
& & & \{4, 8\} & 4 & stable \\
& & & \{5, 10\} & 5 & stable \\
\hline

\end{tabular}
\label{tab:fea_octagon_cases}
\end{table}

\subsection{
General geometries}
\begin{figure}[H]
  \centering
  \includegraphics[width=1.0\linewidth]{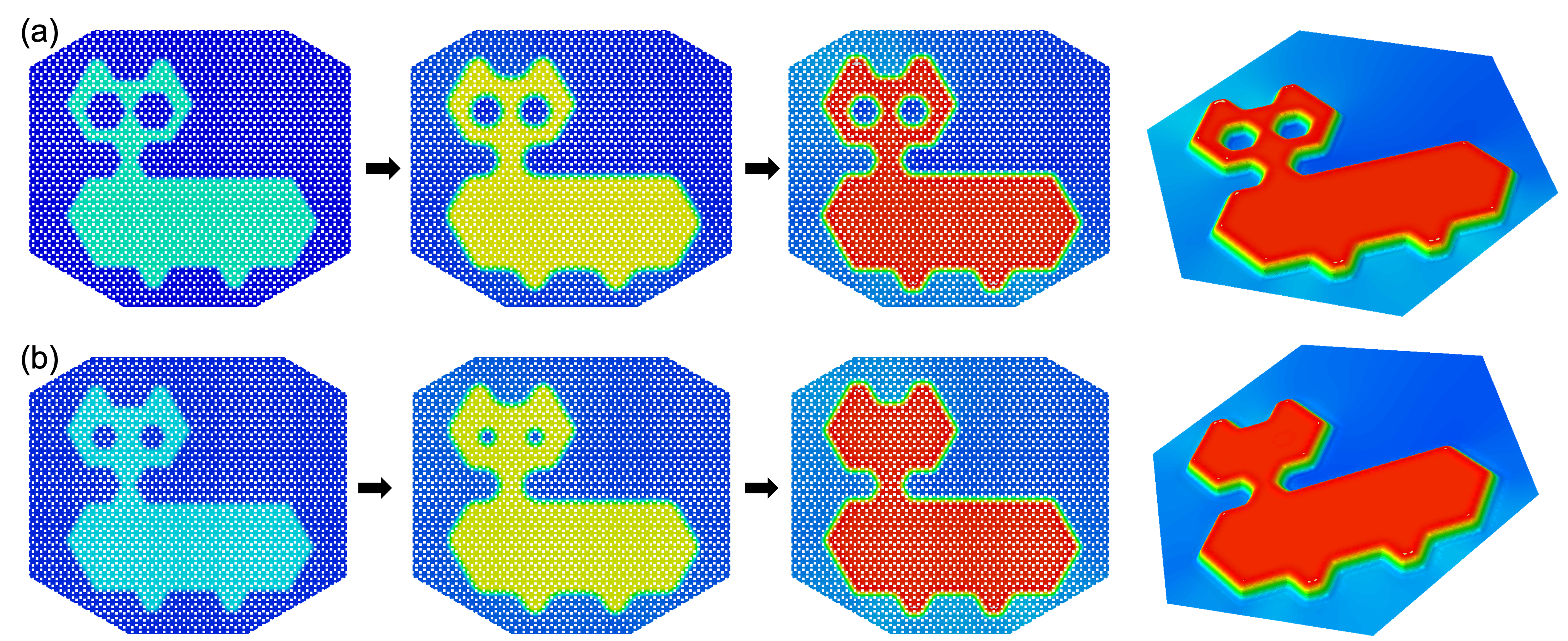}
  \caption{
  A cat-shaped geometry is created by one irregular polygon and two hexagonal domain walls. From left to right, the subfigures illustrate the process of applying the prescribed displacement in the green region to generate the initial domain wall and subsequently releasing the displacement constraint. The color contours represent the out-of-plane displacement $w$, and the color bar is identical to that used in Fig.~\ref{fig1}.}
  \label{cat}
\end{figure}

Having validated the stability inference for both the convex internal angle \(120^\circ\) and concave angle \(240^\circ\), now we are able to create more general geometries. Since with $\alpha=0.50, \gamma=5/16$, the stable range of $n_{\text{st}}$ for $120^\circ$ and $240^\circ$ are the same and almost span all the possible edge lengths (Table~\ref{tab:fea_octagon_cases}),  here we use these parameters in examples. In Fig.~\ref{cat}, three closed domain walls are created to form a cat-like geometry, with the irregular polygon domain wall for the cat's outline and two hexagon domain walls for the cat's eyes. In Fig.~\ref{cat}(b), since the eye domain walls are smaller than the stable range, the eyes evolve inwards and eventually disappear.

\section{Conclusion}\label{section 6}
This work establishes governing principles on the evolution and stability of closed domain walls in an elastic plate supported by a bistable foundation. Through a reduced-order model (ROM) derived via the collective-coordinate method, supported by finite element simulations, we identify the fundamental mechanisms that determine whether a domain wall expands, shrinks, or arrests at a metastable size.

First, for axisymmetric domain walls, we show that three parameters jointly determine the evolution: the bistable asymmetry $\alpha$, the discreteness $\gamma = d/W$, and the domain wall radius $r_{\text{d}}$. In the continuum limit ($\gamma \rightarrow 0$), both FEA and ROM reveal a critical nucleation radius $r_{\text{cr}}(\alpha)$ that separates shrinking and expanding. This behavior arises from the competition between the surface energy, which scales as $O(r_{\text{d}})$, and the bulk energy difference between the two foundation states, which scales as $O(r_{\text{d}}^2)$. The ROM quantitatively predicts $r_{\text{cr}}(\alpha)$ and its rapid increase as $\alpha \rightarrow 0.5^{-}$, in good agreement with FEA.

Second, when the discreteness $\gamma$ exceeds a critical value $\gamma_{\text{cr}}(\alpha)$, the energy landscape is no longer smooth. Instead, periodic Peierls-Nabarro-type (P-N) energy modulations emerge from the discrete foundation, producing local minima in the energy landscape. These minima pin the domain wall and give rise to metastable radii. The ROM correctly predicts (i) that metastability first appears near $r_{\text{cr}}(\alpha)$, the flattest region of the continuum energy landscape; (ii) that $\gamma_{\text{cr}}(\alpha)$ is smallest as $\alpha \rightarrow 0.5^{-}$; and (iii) that the metastable radius only exists in finite range $r_{\text{dst}}(\alpha, \gamma)$ and this range broadens with increasing $\gamma$, arising from the competition between discreteness effects (P-N pinning) and the global energetic drive for phase transition. The ROM accurately captures the lower bound of $r_{\text{dst}}$, and only deviates at large radii, where domain walls in FEA transition from circular to lattice-conforming hexagonal shapes.

Third, we extend the stability analysis to non-axisymmetric geometries. For discrete foundations, the stability of irregular convex polygons with internal angles of $120^\circ$ is well predicted by the stability range of regular hexagons: if every edge length falls within the stable integer range $n_{\text{st}}(\alpha, \gamma)$, the entire polygon is metastable. For polygons including concave $240^\circ$ angles, stability can be inferred by introducing complementary regular hexagons governed by the complementary bistable parameter $\alpha' = 1 - \alpha$. FEA confirms that a configuration is stable only when both the convex and concave edges satisfy their corresponding stability criteria. This framework enables the construction of arbitrary metastable shapes by assembling segments with $120^\circ$ and $240^\circ$ internal angles.

Overall, this study identifies the critical role of discreteness in creating Peierls--Nabarro--type pinning and establishes design rules linking $\alpha$, $\gamma$, and geometric features to shape stability. In the continuous limit, this study provides quantitative guidance on the minimum actuation area required to trigger a global shape change. In the discrete regime, it identifies the foundation density and system parameters necessary to maintain a desired metastable configuration. Beyond the specific plate–bistable-foundation system investigated here, the energy-landscape framework used in this work can be extended to a broader class of multistable metamaterials composed of elastically coupled bistable units. First, the analysis can be generalized to different lattice types, not limited to the hexagonal honeycomb considered here. As long as the discreteness remains within the regime where a smooth axisymmetric domain-wall profile is a reasonable approximation, the procedure of calculating the potential energy and extracting the P–N–type periodic modulation should remain valid for predicting metastable ranges. Second, the analysis can be extended to a broader class of elastic coupling mechanisms and bistable on-site potentials. The role of the plate in mediating coupling can be interpreted through the biharmonic bending operator, which, when discretized, naturally gives rise to next-nearest-neighbor linear elastic interactions. Other forms of linear or nonlinear elastic coupling, as well as alternative bistable potentials, would modify the quantitative predictions, but the qualitative trends associated with the bistable asymmetry and the discreteness are expected to persist.

There are also some limitations on this work:
(i) We assume that the plate–foundation coupling occurs only through the out-of-plane displacement. Consequently, the analysis focuses on bending-dominated deformations and neglects axial strains and large-deformation effects. This approximation may deviate from experimental conditions when the mechanical connection between the plate and the bistable units is non-ideal or when the plate undergoes significant deformation. (ii) In formulating the ROM, we prescribe a fixed tanh-shaped domain-wall profile with a width that does not vary with the domain-wall size or propagation speed. This represents a first-order approximation. As observed in kink motion for the beam–bistable system~(\cite{decker_kink_2021}) and other similar systems (such as $\phi4$ model,~\cite{kevrekidis_dynamics_2000}), the kink width decreases slightly as the propagation speed increases; thus the present approximation is accurate primarily when the domain-wall motion is slow. (iii) The stability inference used to generalize axisymmetric results to irregular geometries is a first-order approximation. Additional FEA studies are needed to further validate, refine, or delimit the applicability of this criterion.

\section*{Declaration of competing interest}
None.

\section*{Data availability}
Data will be made available on request.

\section*{Acknowledgments}
The authors gratefully acknowledge support via the Air Force Office of Scientific Research, award numbers FA9550-23-1-0299 and FA9550-23-1-0416. VT acknowledges partial support from the project grant ExFLEM ANR-21-CE30-0003-01.

\appendix
\section{Derivation of the potential energy}\label{append A}
When the foundation in a continuous limit where $\gamma=d/W \ll 1$. In stage~1 where the domain wall either shrinks or expands, the elastic energy in the plate (\cite{timoshenko1959theory}) is given by~
\begin{equation}\label{plate energy}
\begin{aligned}
E_{p1} &= \frac{1}{2} D \iint \left[\frac{1}{r^2} \left( \frac{\partial w}{\partial r} \right)^2 +\left( \frac{\partial^2 w}{\partial r^2} \right)^2 \right] dA \\
&= \pi D w_b^2\int_0^\infty \left[  \frac{4}{W^2} \text{sech}^4(\frac{r-r_d}{W/4}) \frac{1}{r} +\frac{16^2}{W^4} \text{sech}^4(\frac{r-r_d}{W/4}) \text{tanh}^2(\frac{r-r_d}{W/4}) r \right]  dr \\
&= \pi D w_b^2\int_{-\frac{r_d}{W/4}}^\infty \left[  \frac{4}{W^2} \text{sech}^4(\xi) \frac{1}{\frac{W}{4}\xi+r_d} +\frac{16^2}{W^4} \text{sech}^4(\xi) \text{tanh}^2(\xi) (\frac{W}{4}\xi+r_d) \right] \frac{W}{4} d\xi \\
& \approx \pi D w_b^2\int_{-\infty}^\infty \left[  \frac{4}{W^2} \text{sech}^4(\xi) \frac{1}{r_d} +\frac{16^2}{W^4} \text{sech}^4(\xi) \text{tanh}^2(\xi) \frac{W}{4}\xi +\frac{16^2}{W^4} \text{sech}^4(\xi) \text{tanh}^2(\xi) r_d \right] \frac{W}{4} d\xi\\
& = \pi D w_b^2\int_{-\infty}^\infty \left[  \frac{4}{W^2} \text{sech}^4(\xi) \frac{1}{r_d} +\frac{16^2}{W^4} \text{sech}^4(\xi) \text{tanh}^2(\xi) r_d \right] \frac{W}{4} d\xi\\
\end{aligned}
\end{equation}
where the Poisson ratio is taken as zero for simplicity without losing the key insights by the ROM. In the second line the change of variable $\frac{r-r_d}{W/4}=\xi$ has been used. In stage 1,  $r_d > \frac{W}{2}$ so that $-\frac{r_d}{W/4}<-2$; under this condition, the lower integration limit $-\frac{r_d}{W/4}<-2$ in the third line can be well approximated by $-\infty$ due to the strong localization of $\text{sech}$ function, and the first term in the integration can be approximated well by the firsts term in the fourth line. The second term in the fourth line vanishes as it is odd. Using the identities
\begin{equation}
\int_{-\infty}^\infty \text{sech}^4(\xi) d\xi = \frac{4}{3}, \quad \int_{-\infty}^\infty \text{sech}^4(\xi) \text{tanh}^2(\xi) d\xi = \frac{4}{15}
\end{equation}
yields:
\begin{equation}\label{plate energy result 1}
E_{p1} = \frac{\pi D w_b^2}{15 W^2} \left( 256 \frac{r_d}{W} + 20 \frac{W}{r_d} \right)
\end{equation}
  
The elastic energy stored in the bistable foundation is
\begin{equation}\label{eq:foundation-energy}
\begin{aligned}
E_{f1}
&= \int_{0}^{\infty} 2\pi r\, U\!\big(w(r)\big)\,dr \\
&= 2\pi k \!\int_{0}^{\infty}\!\!\left[ 
\tfrac{1}{4} w^4
-\tfrac{1}{3}(1+\alpha)w_b\,w^3
+\tfrac{1}{2}\alpha w_b^{2}w^{2}
\right] r\,dr \\[4pt]
&= 2\pi k w_b^4 
\!\int_{-\frac{r_d}{W/4}}^{\infty}\!\!\Bigg[
\tfrac{1}{64}\tanh^{4}\xi
+\Big(\tfrac{1}{48}-\tfrac{\alpha}{24}\Big)\tanh^{3}\xi
-\tfrac{1}{32}\tanh^{2}\xi
+\Big(\tfrac{\alpha}{8}-\tfrac{1}{16}\Big)\tanh\xi
+\Big(\tfrac{\alpha}{12}-\tfrac{5}{192}\Big)
\Bigg]\\
&
\Big(r_d+\tfrac{W}{4}\xi\Big)
\tfrac{W}{4}\,d\xi,
\end{aligned}
\end{equation}
where $U(w)$ is the bistable potential Eq.~(\ref{bistable potential}) and the ansatz Eq.~(\ref{tanh1}) is used. In stage 1,  $r_d > \frac{W}{2}$, so that $-\frac{r_d}{W/4}<-2$, allowing the integrations to be well approximated by
\begin{equation}
\begin{aligned}
\int_{-z}^\infty \left ( \text{tanh}\xi-1\right )\xi d\xi \approx z^2-1, 
\quad \int_{-z}^\infty \left ( \text{tanh}^2\xi-1\right )\xi d\xi = 0\\
\quad \int_{-z}^\infty \left ( \text{tanh}^3\xi-1\right )\xi d\xi = 0 \approx z^2-2, 
\quad \int_{-z}^\infty \left ( \text{tanh}^4\xi-1\right )\xi d\xi = 0\\
\int_{-z}^\infty \left ( \text{tanh}\xi-1 \right ) d\xi \approx -2z, 
\quad \int_{-z}^\infty \left ( \text{tanh}^2\xi-1 \right ) d\xi = -2\\
\quad \int_{-z}^\infty \left (  \text{tanh}^3\xi-1 \right ) d\xi \approx -2z, 
\quad \int_{-z}^\infty \left ( \text{tanh}^4\xi-1 \right ) d\xi \approx -\frac{8}{3},
\end{aligned}
\end{equation}
It can be shown that Eq.~(\ref{eq:foundation-energy}) can be rearranged as a linear combination of the above integrations, and thus it yields to:
\begin{equation}\label{foundation energy result 1}
E_{f1} 
\approx   \frac{\pi k\, w_b^{4} W^{2}}{384}
\left[
   (2\alpha - 1)
   \;+\;
   4\,\frac{r_d}{W}
   \;+\;
   32\,(2\alpha - 1)\left(\frac{r_d}{W}\right)^{2}
\right]
\end{equation}

Next, we consider stage 2 where the domain wall disappears. The elastic energy in the plate in stage 2 follows from Eq.~(\ref{plate energy result 1}) with $w_b$ replaced by $H(t)$ and $W(t)=2r_d(t)$:
\begin{equation}\label{plate energy result 2}
E_{p2} = \frac{14\pi D}{5} \frac{H^2(t)}{{r_d(t)}^2}
\end{equation}

The elastic energy in the foundation is:
\begin{equation}\label{foundation energy 2}
\begin{aligned}
E_{f2} &= \int_0^\infty 2\pi r \, U(w(r)) \, dr \\
&= 2\pi  k \int_0^\infty \left [ \frac{1}{4}w^4-\frac{1}{3}(1+\alpha)w_bw^3 +\frac{1}{2}\alpha w_b^2  w^2\right ]   r dr \\
&\approx \frac{\pi k w_b^4}{96}  \left [ 11\,{\left ( \frac{H(t)}{w_b} \right )}^4-(18\alpha+18)\,{\left ( \frac{H(t)}{w_b} \right )}^3+36 \alpha\,{\left ( \frac{H(t)}{w_b} \right )}^2 \right ] {r_d(t)}^2
\end{aligned}
\end{equation}
which recovers Eq.~(\ref{foundation energy result 1}) for $H = w_b$ and $r_d = W/2$.

\section{Derivation of the kinetic energy} \label{append B}
The kinetic energy in stage 1 can be calculated form the velocity field Eq.~(\ref{velocity}):
\begin{equation}\label{kinetic}
\begin{aligned}
E_{k1} &= \iint \frac{1}{2} \rho h v^2 \, dA \\
  &= \int_0^\infty \frac{1}{2} \rho h v^2 \cdot 2\pi r \, dr \\
  &= \frac{4\pi \rho h w_b^2}{W^2} \dot{r}_d^2 \int_0^\infty \text{sech}^4\left( \frac{r - r_d}{W/4} \right) r \, dr \\
\end{aligned}
\end{equation}
For this integration, a change of variable $\frac{r-r_d}{W/4}=\xi$ yields
\begin{equation}
\begin{aligned}
  \int_0^\infty \text{sech}^4\left( \frac{r - r_d}{W/4} \right) r \, dr &= \int_{-\frac{r_d}{W/4}}^\infty \text{sech}^4(\xi) \left( \frac{W}{4} \xi + r_d \right) \frac{W}{4} d\xi\\
  &= \frac{W^2}{16} \int_{-\frac{r_d}{W/4}}^\infty \text{sech}^4(\xi) \xi \, d\xi + \frac{W}{4} r_d \int_{-\frac{r_d}{W/4}}^\infty \text{sech}^4(\xi) d\xi
\end{aligned}
\end{equation}
Due to  $-\frac{r_d}{W/4}<-2$, the  lower integration limit can be well approximated by $-\infty$ due to the localized property of $\text{sech}$ function. The first term of the integration vanishes since $\text{sech}^4(\xi) \xi$ is odd, and the second term evaluates to 
\begin{equation}
  \int_{-\infty}^\infty \text{sech}^4(\xi) d\xi = \frac{4}{3}
\end{equation}
Therefore, the kinetic energy in Eq.~(\ref{kinetic}) becomes  
\begin{equation}\label{kinetic_result}
E_{k1} = \frac{4\pi \rho h w_b^2 }{3} \frac{r_d}{W} \dot{r}_d^2 
\end{equation}

\bibliographystyle{elsarticle-harv} 
\bibliography{References}

\end{document}